\documentclass[journal]{IEEEtran}
\usepackage{multirow}
\usepackage{amsmath,amssymb,amsfonts}
\usepackage{algorithmic}
\usepackage{graphicx}
\usepackage{textcomp}
\usepackage{xcolor}
\usepackage{amsmath,amssymb,amsfonts}
\usepackage{colortbl}
\usepackage{caption}
\usepackage{subcaption}
\usepackage{wrapfig}
\usepackage[utf8]{inputenc}
\usepackage{mathtools}
\usepackage{breqn}
\usepackage{caption}
\usepackage{subcaption}
\usepackage{tabularx}
\usepackage{authblk}
\usepackage{color, colortbl}
\usepackage{bbding}
\usepackage{pifont}
\usepackage{wasysym}
\usepackage{amssymb}
\usepackage{array}

 \usepackage{array,multirow,graphicx}
 \usepackage{float}

 \newcommand*\rot{\rotatebox{90}}

\definecolor{Gray}{gray}{0.9}

\newcommand{\etal}{\textit{et al. }}
    \usepackage{enumitem}
\newcommand{\tablistcommand}{
            \leavevmode\par\vspace{-\baselineskip}
                            }
\newlist{tableitems}{itemize}{1}
\setlist[tableitems]{nosep,     
                     topsep     = 0pt               ,
                     partopsep  = 0pt               ,
                     leftmargin = *                 ,
                     label      = \textbf{-}        ,
                     before     = \tablistcommand   ,
                     after      = \tablistcommand
                     }
\ifCLASSINFOpdf

\else

\fi

\begin{document}
%
\title{Understanding the Trustworthiness Management in the Social Internet of Things: A Survey}

\author{Subhash~Sagar,
        Adnan~Mahmood,
        Quan~Z.~Sheng,
        Jitander~Kumar~Pabani,
        and~Wei~Emma~Zhang
\thanks{Subhash Sagar, Adnan Mahmood, and Quan Z. Sheng are with the School of Computing, Macquarie University, Sydney,
NSW 2109, Australia, e-mail: \{subhash.sagar, adnan.mahmood, michael.sheng\}@mq.edu.au.}
\thanks{Jitander Kumar is with the School of Telecommunication Engineering, University of Malaga, Malaga, 29016, Spain, e-mail: jitander.pabani@uma.es.}
\thanks{Wei Emma Zhang is with the School of Computer Science, Faculty of Engineering, Computer, and Mathematical Sciences, The University of Adelaide, Adelaide, SA 5005, Australia, e-mail: wei.e.zhang@adelaide.edu.au.}
}


\maketitle

\begin{abstract}
The next generation of the Internet of Things (IoT) facilitates the integration of the notion of social networking into smart objects (i.e., things) in a bid to establish the social network of interconnected objects. This integration has led to the evolution of a promising and emerging paradigm of the Social Internet of Things (SIoT), wherein the smart objects act as \emph{social objects} and intelligently impersonate the social behaviour similar to that of humans. These social objects are capable of establishing social relationships with the other objects in the network and can utilize these relationships for service discovery. Trust plays a significant role to achieve the common goal of trustworthy collaboration and cooperation among the objects and provide systems’ credibility and reliability. In SIoT, an untrustworthy object can disrupt the basic functionality of a service by delivering malicious messages and adversely affect the quality and reliability of the service. In this survey, we present a holistic review of trustworthiness management for SIoT. The essence of trust in various disciplines has been discussed along with the Trust in SIoT followed by a detailed study on trust management components in SIoT. Furthermore, we analyze and compare the trust management schemes by primarily categorizing them into four groups in terms of their strengths, limitations, trust management components employed in each of the referred trust management schemes, and the performance of these studies vis-à-vis numerous trust evaluation dimensions. Finally, we 
discuss the future research directions of the emerging paradigm of SIoT particularly for trustworthiness management in SIoT.    
\end{abstract}
%
\IEEEpeerreviewmaketitle
\section{Introduction}
%
%
%
%
\IEEEPARstart{T}{he} notion of the Internet of Things (IoT) was prophezied by Kevin Ashton \cite{ashton2009internet} in 1999 as a key paradigm, wherein humans and devices, i.e., objects, would connect and interact over the Internet. Over the last decade or so, this technological viewpoint of IoT became a reality since a network of billions of smart objects (often also referred to as the \emph{`things'}) began connecting over the Internet. This evolution of connected smart objects has, therefore, contributed to the considerable number of applications and services having practical implications in our daily lives \cite{ATZORI20102787}\cite{GUBBI20131645}\cite{10.1145/3464960}\cite{WoT-book}. 
Some of such applications and services 
fall 
in the domains of healthcare, smart cities, smart homes, and smart agriculture. In terms of healthcare, there are several applications, e.g., telemedicine to facilitate doctors to monitor the health of patients via wearables embedded with IoT, clinical analytics to study patients together with the site performance, and mhealth to provide two-way communication between the doctors and the patients by employing personal devices \cite{Yang2020}\cite{8993839}. 
Smart cities can benefit from a variety of applications too, e.g., smart grid and smart energy systems can be employed for energy-saving purposes via monitoring of power utilization to optimize energy cost (so as to provide a better consumer service), and fault detection due to environmental hazards; smart transportation system to reduce the travel time and for ensuring efficient traffic management to mitigate the traffic congestion; and smart waste management for facilitating the cities' administration to efficaciously manage and handle massive and ever-increasing volumes of municipal waste via installing of smart bins \cite{8030479}\cite{8675165}. 
Smart home applications include smart home automation, smart lighting, smart doors (and windows), and smart kitchen appliances \cite{ALAA201748}\cite{zaidan2020review}. 
Finally, smart agriculture can strengthen the traditional farming via precision farming to control and manage the livestock and crops more accurately via the agriculture drones, livestock monitoring, and smart greenhouses \cite{9316211}\cite{PATHAN202081}.

The development of these applications and services are realized via smart physical objects, 
e.g., 
a variety of sensors and actuators, radio-frequency identification devices (RFIDs) \cite{ShengLZ08}, smartphones, other data processing devices, 
possessing the capability to collect, monitor, and analyze the data pertinent to human life and usually interact with one another to accomplish a common goal. This survey interchangeably uses the terminologies of \emph{objects}, \emph{nodes} and \emph{device} to refer to the IoT-enabled things, and trust models are also referred to as trust management systems or trustworthiness management systems.
%
%
\renewcommand{\arraystretch}{1.25}
\newcolumntype{C}[1]{>{\centering\arraybackslash}p{#1}}
\newcommand{\centered}[1]{\begin{tabular}{l} #1 \end{tabular}}
\begin{table*}[ht]
\footnotesize
\setlength{\arrayrulewidth}{.1em}
  \begin{center}
    \caption{Comparison with recent surveys}
    \label{tab:survey}
    \begin{tabular}{ |>{\centering\arraybackslash}m{2cm} |>{\centering\arraybackslash}m{0.9cm} |>{\centering\arraybackslash}m{0.9cm} |>{\centering\arraybackslash}m{0.9cm} |>{\centering\arraybackslash}m{0.95cm}|>{\centering\arraybackslash}m{0.9cm}| >{\centering\arraybackslash}m{8.4cm}|}
     \hline
     \rowcolor{Gray}
      \small\textbf{Survey} & \small\textbf{TM-C} & \small\textbf{TM-S} & \small\textbf{TS-A} & \small\textbf{SIoT-P} & \small\textbf{TM-R} & \small\textbf{Description} \\ 
      \hline 
        \small Abdelghani \etal\textbf{\cite{10.1007/978-3-319-45234-0_39}} & \small$\sim$ & \small$\sim$ & \small\textbf{\ding{53}} & \small\textbf{\ding{53}} & \small\textbf{\ding{53}} & \begin{tableitems}[nosep,after=\strut]
                        \item This study presents the comparative analysis of the trust management model for the SIoT environment by taking into account the trust properties and SIoT constraints. \vspace{-1em}
        \end{tableitems} \\ \hline
        \small Rashmi \etal\textbf{\cite{10.1007/978-981-13-5802-9_19}} & \small\textbf{\ding{53}} & \small$\sim$ & \small\textbf{\ding{53}} & \small\textbf{\ding{53}} & \small\textbf{\ding{53}} & \begin{tableitems}[nosep,after=\strut]
                        \item This study on trust management in SIoT delineates an overview of trust management studies in SIoT and compared the same with different performance metrics and trust-related attacks. \vspace{-1em}
        \end{tableitems} \\ \hline
        \small Amin \etal\textbf{\cite{app9010166}} & \small$\sim$ & \small\textbf{\ding{53}} & \small\textbf{\ding{53}} & \small\textbf{\ding{53}} & \small$\sim$ & \begin{tableitems}[nosep,after=\strut]
                        \item The survey discusses the trust and friendliness-based approaches in terms of scalability, adaptability, and network structure by taking into account the aspects of service composition and social similarity. \vspace{-1em}       \end{tableitems} \\ \hline
        \small Roopa \etal\textbf{\cite{MS201932}} & \small$\sim$ & \small\textbf{\ding{53}} & \small\textbf{\ding{53}} & \small\textbf{\ding{53}} & \small$\sim$ & \begin{tableitems}[nosep,after=\strut]
                        \item This study provides a comprehensive overview of current research trends in the SIoT paradigm. Service discovery and composition, relationship management, network navigability, and trustworthiness management are among the discussed trends.\vspace{-1em}  
        \end{tableitems} \\ \hline
        \small Chahal \etal\textbf{\cite{CHAHAL202013}} & \small\textbf{\ding{51}} & \small$\sim$ & \small\textbf{\ding{53}} & \small\textbf{\ding{53}} & \small$\sim$ & \begin{tableitems}[nosep,after=\strut]
                        \item This survey presents a detailed comparison of protocols, architectures, and trust management for SIoT where the most emphasis is given to trust management components employed in the literature. \vspace{-1em}   
        \end{tableitems} \\ \hline
        \small Khan \etal\textbf{\cite{9264256}} & \small\textbf{$\sim$} & \small\textbf{$\sim$} & \small\textbf{\ding{53}} & \small\textbf{\ding{53}} & \small\textbf{$\sim$} & \begin{tableitems}[nosep,after=\strut]
                        \item This survey discusses a comparative and comprehensive analysis of SIoT architecture, trust management systems, and open research challenges in SIoT..\vspace{-1em}  
        \end{tableitems} \\ \hline
        \small\leavevmode\color{black} This Survey & \small\textbf{\ding{51}} & \small\textbf{\ding{51}} & \small\textbf{\ding{51}} & \small\textbf{\ding{51}} & \small\textbf{\ding{51}} & \begin{tableitems}[nosep,after=\strut]
                        \item This survey provides an extensive study on trust management components, a comparison of trust management schemes, an overview of trust in SIoT-based applications and SIoT platforms, and a summary on future research direction on trust management in SIoT. \vspace{-1em}    
        \end{tableitems} \\  \hline
        \multicolumn{7}{|c|}{\text{Fully Covered: \textbf{\ding{51}}, Not Covered: \textbf{\ding{53}}, Partially Covered: \textbf{$\sim$}}}\\ 
        \multicolumn{7}{|c|}{\text{\textbf{TM-C}: Trust Management Components, \textbf{TM-S}: Trust Management Schemes, \textbf{TS-A}: Trust in SIoT-based Applications}}\\ 
        \multicolumn{7}{|c|}{\text{\textbf{SIoT-P}: SIoT Platform, \textbf{TM-R}: Trust Management Research Challenges}}\\ \hline
    \end{tabular}
  \end{center}
  \vspace{-3mm}
\end{table*}

With the advancement in the IoT applications, it is anticipated that there would be around 75 billion interconnected devices worldwide by 
2025 \cite{iot-devices} and international data corporation gives the worldwide IoT spending estimation of about \$742 billion in 
2020, and expects to achieve a growth rate of 11.3\% in the period from 2021 to 2024 \cite{torchia2017worldwide}. Furthermore, IoT is foreseen to have a considerable financial impact of upto \$11.1 trillion on the global economy by 2025, wherein factories operations and equipment optimization will have the highest growth of around \$3.7 trillion followed by retail environment, logistics and navigation, smart cities (i.e., public health and transportation), autonomous vehicles, etc, \cite{manyika2015internet}. Moreover, these billions of IoT devices result in a substantial amount of data exchange, and for which a state-of-the-art networking infrastructure is highly indispensable to not only reveal the undiscovered operational efficiencies and devise an end-to-end ecosystem incorporating individuals' needs \cite{ATZORI20102787}. In short, IoT is described as a dynamic and a global network of infrastructure, emphasising physical and virtual objects with the capability to collect the human and environmental characteristics supporting interoperability using intelligent interfaces and standard communication protocols \cite{GUBBI20131645}\cite{9011598}\cite{van2008internet}. 

\subsection{From IoT to Social Internet of Things (SIoT)}
As IoT is of great benefit in various applications, 
numerous challenges including but not limited to heterogeneity, service discovery and composition, and scalability necessitate for designing and developing the IoT infrastructure \cite{9319033}\cite{mukhopadhyay2014internet}\cite{sisinni2018industrial}\cite{hassan2017internet}. Heterogeneity is of the main concern since an IoT network comprises of several devices each of varying nature and manufacturer specific operating systems and protocols. This heterogeneous nature impedes the common solution for application development and thus the system needs a shared communication paradigm among the devices. Furthermore, information and service discovery is another challenge that needs a novel trusted protocol to ease the exploitation of trust-related services, and with the enormous number of objects, existing solutions to these problems do not scale up. Therefore, a possible way is to adopt the human sociological behaviour to scale up the current solution. It is pertinent to mention that humans themselves are heterogeneous, complex, and dynamic in nature, nevertheless, there still exists the notion of social relationship that facilitates in forming the societies among humans based on common interests and needs. Subsequently, the information discovery in humans is possible through the principle of small-world phenomena originally suggested by Jon Kleinberg that refers to the short chain of links among the individuals in societies  \cite{Kleinberg2011}\cite{10.5555/2980539.2980596}. Over the last decade or so, in view of human societies, there has been a lot of research endeavors by scientists in academia and industry, analyzing the possibilities of integrating the paradigm of social networking into the IoT ecosystem \cite{Kranz2010ThingsTT}\cite{guinard2010sharing}\cite{5722081}.  Moreover, Holmquist \etal \cite{10.1007/3-540-45427-6_10} introduced the idea of socialization amongst the objects, wherein an easy-to-use technique was proposed to establish the relationship between the objects via utilizing the context proximity.   
\textbf{\begin{figure*}[t]
    \centering
    \includegraphics[width=\linewidth]{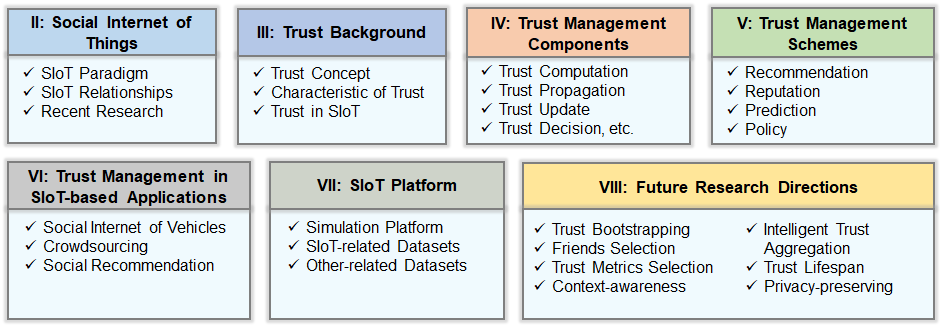}
    \caption{Taxonomy of this survey}
    \label{fig:survey_tax}
    \vspace{-1em}
\end{figure*}}

The emerging paradigm of SIoT employs the said integrate concept, wherein each object is not only capable of capturing surrounding characteristics but is also able to establish the relationship with the other objects in the network. The enhanced capabilities (i.e., socializing with other objects) of these intelligent social objects results in efficient collaboration as they establish their own social network to manage the social relationships and social communities in order to provide intelligent decision making without human intervention \cite{Atzori2011SIoTGA}\cite{ATZORI20123594}. In light of the fact that SIoT can benefit in numerous research gaps including but not limited to the efficient discovery of services and objects, to ensure the scalability like in human social network, managing social relationships among the intelligent social objects, network navigability with the idea of smart-world phenomena whereby utilizing the relationships among the objects, and trustworthiness management among the participating smart objects \cite{MS201932}\cite{nitti2014_n}. 

Moreover, the social characteristics in SIoT have paved the way for the next generation of IoT in a bid to discover the required services via utilizing the social relationships with the neighbouring objects. Nevertheless, the risks and uncertainty may diminish the significance of SIoT paradigm primarily owing to the challenges pertinent to security, privacy and trust of these intelligent social objects \cite{7576667}. For instance, when an object request for a specific service (referred to as a service requester), then, different service providers may acknowledge the same in order to  to provide the requisite service and this is where the trustworthiness of these service provider come into play since the one possessing the highest trust would be opted for the requisite service. Besides, security and privacy plays an important role for deploying and commercializing of the SIoT services. Although traditional solutions i.e., cryptographic and non-cryptographic ones have been proposed to tackle such challenges \cite{9011598}\cite{9382796}, nevertheless, security challenges like trust and/or reputation are difficult to get addressed via such solutions. Likewise, there exist malicious objects that can disrupt the basic functionality of a network for malicious purposes by damaging the reputation of good (well behavioured) objects or by increasing the trustworthiness of misbehaving objects \cite{Fan2020}\cite{9328204}. An efficient trust management system in SIoT is, therefore, imperative for dealing with the misbehaving objects (which are capable of jeopardizing the network functionality) by restricting the services of such nodes and via selecting the reliable and trustworthy objects before relying on the information provided by them.          

\subsection{Existing Surveys on Trust Management in SIoT}
To date, a plethora of surveys on trust management for IoT \cite{AHMED2019102409}\cite{9043531}\cite{SHARMA2020475} have been presented in the research literature, however, there are only a few surveys that offer a detailed insight on trust management systems for the SIoT paradigm. In 2016, Abdelghani \etal \cite{10.1007/978-3-319-45234-0_39} published the first survey on trust management in SIoT that briefly discussed the SIoT concept, its trust properties, and compared the presented trust models in terms of varying dimensions, i.e., scalability, adaptability, and resiliency, Nevertheless, this survey lacks a comprehensive discussion on trust management systems and the trust management components employed in the presented studies. A recently published survey \cite{10.1007/978-981-13-5802-9_19} on trust management in SIoT provides an overview of trust management studies in SIoT and compared the same in terms of different performance metrics, e.g., scalability, adaptability, power efficiency, survivability, and resiliency.
However, this survey still lacks reviewing many important aspects of trust, including but not limited to trust components, recent studies, and open research challenges. Furthermore, Amin \etal \cite{app9010166} published a survey on trust and friendliness approaches for SIoT, wherein the notion of SIoT is reviewed in view of enabling technologies i.e., clouds, multiagent, and Industry 4.0, followed by a comparison of different approaches of trust and friendliness in SIoT. Nevertheless, the analysis on trust management schemes, particularly for SIoT, are not discussed. 

A holistic view of the SIoT paradigm is explored in \cite{MS201932}, wherein the current research trends in SIoT, i.e., service discovery and composition, relationship management, network navigability, and trustworthiness management are 
investigated. Yet, this survey still lacks the comparison of the latest trust management schemes in the SIoT paradigm as it encompasses the discussion on subjective/objective and dynamic trust management schemes. One of the comprehensive survey on trust management in SIoT is published by Rajanpreet \etal \cite{CHAHAL202013}. This survey compares and analyzes the trust management system in multiple domains, i.e., wireless sensor networks and IoT, and subsequently presents the detailed explanation of trust management components employed in the literature. Yet, the comparison of is not solely based on trust management schemes in SIoT but it also includes trust management in IoT, and the clarification of current research challenges for trust computation is also not discussed. 
The most recent survey on trust management in SIoT is published by Wazir \etal \cite{9264256} in 2021, wherein the similarities between the IoT and SIoT domains are clarified; SIoT related architectures are comprehensively discussed; and the trust management system for SIoT are comparatively analyzed along with the discussion on future research challenges in SIoT. In view of trust management in SIoT, this study lacks the analysis of trust in SIoT-based applications, discussions on SIoT platforms, and the future research challenges especially in terms of trust computation. Overall, Table \ref{tab:survey} summarizes the researched surveys on trust management in SIoT and also discusses the enhancement in our survey.

\subsection{Main Contributions of This Survey}
To address the aforementioned shortcomings in the existing body of literature, this survey targets the topics and approaches which have not yet been covered. Furthermore, the convenience of readers is kept in mind in order to to present this survey in a way that is self-sufficient by including fundamentals of SIoT, the notion of trust, and trust management components in SIoT. After identifying the significance of a trust management system, this survey entails a comprehensive review of trust management schemes in the existing body of literature. The main contributions of our survey are as follows: 

\begin{itemize}
\item[1)] We deliberate the SIoT paradigm and current research trends in SIoT, the fundamentals of \textit{trust} in various disciplines and the trust management components in SIoT;
\item[2)] Subsequently, we distinguish the trust management perspective and categorize the trust management systems into four broad schemes. In particular, a comparative analysis of these schemes in terms of strengths and limitations is discussed. Moreover, a detailed analysis of these schemes is also performed on the basis of trust evaluation parameters;
\item[3)] We review the trust in three SIoT-based applications with their respective research challenges, summarize the SIoT platform used in the literature for simulation purposes, and also discuss the datasets currently employed for performance evaluation of trust management solutions;
\item[4)] We present a generalized trustworthiness management framework for SIoT that considers the holistic view of trust management process employed for SIoT in the studied literature; and
\item[5)] We identify the future research directions for trustworthiness management in SIoT, particularly, for trust computation purposes.  
\end{itemize}

As a whole, this paper presents a comprehensive review on the recent advancements in trustworthiness management in SIoT and provides a way forward for future research directions. A taxonomy of this survey is depicted in Figure \ref{fig:survey_tax}.

\subsection{Paper Selection}
The articles selected in this paper are high quality papers from reputed transactions (e.g., IEEE Transactions on Knowledge and Data Engineering, IEEE Transactions on Information Forensics and Security, IEEE Transactions on Dependable and Secure Computing, etc.), journals (e.g., Internet of Things, Computer Networks, etc.), and conferences including but not limited to INFOCOM, ICDCS, and PerCom. At first, the articles' selection process involved the search strings such as “trustworthiness” or “trust” or “trustworthy” + “social internet of things” or “SIoT” or “Social IoT” from resource libraries like IEEE, ACM, Elsevier, Springer, Google Scholar, etc. Successively, the articles are further categorized in terms of top journals and conferences. Moreover, we have included the early access papers from these libraries as well as from arXiv\footnote{https://arxiv.org/}. Finally, the papers are selected based on quality, method novelty, employed social trust metrics, and the proposed techniques that directly influence the scope of this paper. 

The remainder of this paper is organized as follows. 
Section~\ref{sec:siot} delineates the SIoT paradigm and current research trends in SIoT. Section~\ref{sec:backg} deliberates the concept of trust in various disciplines, i.e., sociology, psychology, economics, computer science and in SIoT. In Section~\ref{sec:tmc}, trust management components are discussed in detail. Section~\ref{sec:TMS} presents the comparative analysis of the current state-of--the-art trust managements schemes. Section~\ref{sec:tsa} discusses the trust in SIoT-based applications and a number of 
SIoT-platforms along with SIoT related datasets are briefly discussed in Section~\ref{sec:siot_data}. Finally, Section~\ref{sec:siot_future} provides the future research directions for trustworthiness management in SIoT, whereas, concluding remarks are presented in Section~\ref{sec:conc}.

\section{Social Internet of Things (SI\MakeLowercase{o}T)}
\label{sec:siot}
This section provides a fundamental concept of the SIoT paradigm and its significance in terms of various social relationships, and recent research activities and advancements in SIoT.
\begin{figure}[!bt]
    \centering
    \includegraphics[width=\linewidth]{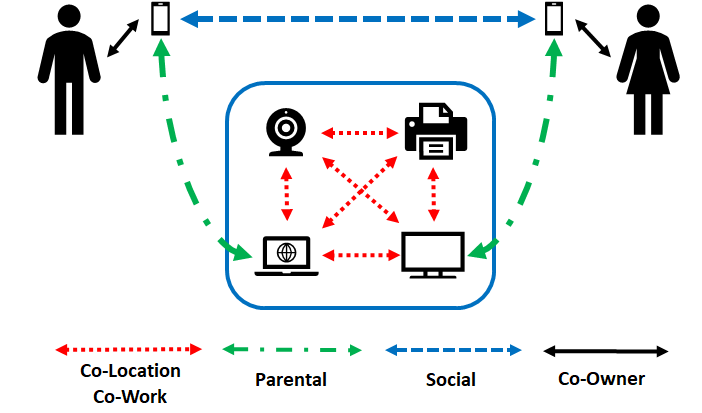}
    \caption{Types of relationships in SIoT}
    \label{fig:siot_rel}
     \vspace{-1em}
\end{figure}
\subsection{SIoT Paradigm}
%
The idea of socialization of objects first conceived in 2001 by P. Mendes \cite{Mendes2011SocialdrivenIO} wherein the idea of objects participating in the conversation similar to the human social network is presented. Similarly, the authors in \cite{guinard2010sharing} explored certain scenarios wherein a person with a smart object can share a particular service with their friend's smart objects through their social circle before the formalization of the SIoT concept by L. Atzori \etal \cite{Atzori2011SIoTGA}. Subsequently, the concept of the SIoT paradigm has emerged which is intended as, \textit{``the integration of social networking concepts into the IoT domain, wherein each object (referred to as Social Object) is capable of establishing social relationships autonomously with the other objects in the network as per the rules and policies set by their respective owners''}. The SIoT characteristics are highly dependent on social relationships among the objects (Figure \ref{fig:siot_rel}) and owners of the objects and some of the frequently occurred relationships are:
%
%
\subsubsection{Ownership Object Relationships (OOR)} OOR represents the relationships between the objects and their respective owners, i.e., an owner can have multiple devices like smartphones, tablets, laptops, etc. This type of relationship results in a high probability of interaction \cite{ATZORI20123594}.
\subsubsection{Social Object Relationships (SOR)} Similar to humans, this type of relationship is established when two or more objects come in contact with each other. For instance, if two individuals are friends and they meet each other regularly then their smartphones may establish a social relationship based on the rules and policies set by their owners \cite{ATZORI20123594}. 
\subsubsection{Parental Object Relationships (POR)} POR is correlated with similar objects having the same manufacturer and same production batch within a given period of time. For example, two smartphones of the same model and same manufacturer may establish this type of relationship \cite{ATZORI20123594}. 
\subsubsection{Co-location Object Relationships (CLOR)} CLOR represent the relationship among the objects possessing the same location e.g. if two or more objects (e.g., sensors and actuators) provide the services in a home or in a industrial automation environment. 
\subsubsection{Co-work Object Relationships (CWOR)} In contrast to CLOR, CWOR signify the relationship involving two or more objects collaborate with each other in a common IoT application in order to accomplish a shared goal. The emphasis in CWOR is on the working relationships between the objects rather than their locations \cite{ATZORI20123594}. 

There are a few unpopular relationships, such as sibling object relationships, guest object relationships, guardian object relationships, stranger object relationships, and service object relationships 
\cite{MS201932}.

Furthermore, the SIoT paradigm conveys numerous desirable implications into a future world populated by intelligent objects encompassing the daily life of human beings and aims to support many applications and services by effectively enhancing the service discovery and composition. Moreover, applying social networking concepts to the IoT unquestionably prompts favorable circumstances that stretch (i) from the enhanced viability, scalability, and the prompt navigability of the network with billions of objects that will populate the IoT in the future (ii) to the arrangement of a degree of trustworthiness that can be built up by utilizing the social relationships among things that are friends and/or having similar interests, and (iii) to the interoperability between the heterogeneous objects. This can be accomplished by exploring the social network and utilizing trustworthy relationships with companion objects.

\subsection{Recent Research Activities in SIoT}
\begin{figure}[t]
    \centering
    \includegraphics[width=0.95\linewidth]{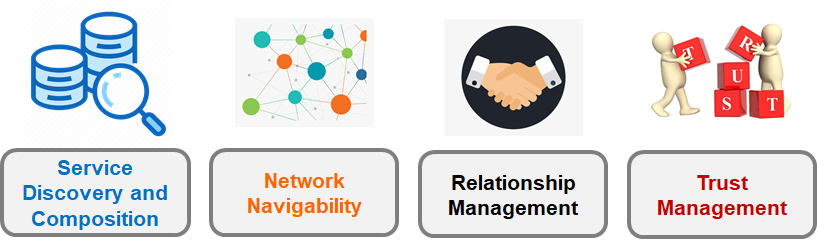}
    \caption{Current research areas in SIoT}
    \label{fig:siot_areas}
    \vspace{-1em}
\end{figure}

In recent years, numerous research articles have been published 
which provide a detailed insight into the SIoT paradigm and its architectures \cite{6962156} \cite{article2}. The authors in \cite{Atzori2011SIoTGA}\cite{ATZORI20123594} introduced the idea of integrating social networking concepts into the IoT in a bid to cope with the issues of service discovery and composition. Besides, the suggested paradigm further facilitates in understanding how an IoT object can establish and manage social relationships with the other objects in a given network. Hence, the resulting paradigm, i.e.,  SIoT, can support novel applications and services for the IoT systems in an efficient and effective manner. Moreover, various SIoT related challenges are studied in the research literature and Figure 
\ref{fig:siot_areas}
 depicts some of these current open key challenges in the 
 SIoT landscape, including but not limited to, i) service discovery and composition \cite{Li2020}\cite{5951906}, ii) network navigability \cite{abdul2018exploiting}\cite{Amin2019}, iii) relationship management \cite{ATZORI20123594}\cite{6994231}, and iv) trust management \cite{MS201932}\cite{Nitti2014}. 

\subsubsection{Service Discovery and Composition} The underlying rationale behind the IoT and SIoT paradigm is to provide the services 
(e.g., 
healthcare, agriculture monitoring)
to the end users. Accordingly, service discovery is of considerable significance and is aimed at discovering the objects and their offered services within real-time environments \cite{8334540}. As the number of devices are increasing at an unprecedented pace, so as the data exchange between them. It is pertinent to mention that the the generated data from these devices is not useful for everyone, and therefore, service discovery is imperative for searching the smart objects providing the useful information in a highly dynamic environment. SIoT facilitates at discovering the service, i.e.,  similar to the humans searching for information in their social network by employing different relationships, hereby providing a scalable solution for service discovery. Subsequently, the service composition provides and enables the interaction between the smart objects subsequent to service discovery \cite{Hussein2017}\cite{9181080}.  

\subsubsection{Network Navigability} To make the service discovery process more efficient by utilizing various relationships (e.g., 
friendship, communities, location)
and use these social link to navigate the network, thus reducing the average path length between the participating objects (i.e., service requester and service provider) \cite{abdul2018exploiting}. In SIoT systems, an object utilizes friends and its friends of friends to search a specific service, nevertheless, it is not feasible for an object to establish a relationship (i.e., to make friendship) with all the objects, and accordingly, a number of researchers have proposed the idea of employing friendship selection methods for choosing minimal friends and to provide the network navigability with reduced path length between the pair of objects using the friendship links \cite{nitti2014_n}\cite{Amin2019}\cite{MOHAMMADI2021107479}.  
\subsubsection{Relationship Management} Relationship management provides the way of embedding the intelligence into smart objects, to make them recognize the friends and foes, and to originate, update and terminate the relationship. The authors in \cite{zhang2012cognitive} introduced the notion of cognitive IoT to integrate the intelligence in IoT objects, wherein their goal was to enable the objects to perceive and sense the physical world. However, the SIoT paradigm requires the objects to recognize not only the physical world but also the social world, and this integration demands further exploration \cite{ATZORI20123594}. Many research efforts have been made over the years to provide the novel ideas for relationship management in terms of friendship selection in the SIoT landscape, wherein different genetic algorithms and appropriate policies have been proposed \cite{6994231}\cite{8273022}\cite{10.1007/978-3-030-49435-3_7}. 
\subsubsection{Trustworthiness Management} The notion of trust ensures reliable and trustworthy interactions by employing trustworthy social relationships among the objects. A plethora of trustworthiness management systems have been proposed in the literature and have been widely employed in various disciplines (e.g., 
sociology, psychology, economics, and computer science \cite{hardin2002}\cite{schlenker1973effects}\cite{doi:10.1177/002224299405800302}), and numerous applications (e.g., 
IoT \cite{AHMED2019102409}, Internet of Vehicles (IoV) \cite{mahmood2019hybrid}\cite{articleadnan}, mobile and vehicular ad-hoc networks \cite{7312406}\cite{9025055}, peer-to-peer networks \cite{7407618}, online social networks \cite{9142365}, e-Commerce \cite{8618376}). Nevertheless, the SIoT paradigm requires the trust management systems that not only deals with the objects but also the social relationships among them. Thus the techniques proposed in the literature cannot be applied directly in the SIoT environment \cite{Nitti2014}. Recently, numerous studies have been published on trust management in the SIoT environment and a comparative analysis of the same has been summarized in Table \ref{tab:recommendation}-\ref{tab:policy} by highlighting their respective strengths and limitations. Moreover, Table \ref{tab:trust_components} illustrates the trust management components employed in these studies, whereas Table \ref{tab:comp_const} delineates the evaluation of these studies with various dimensions.

\section{Background}
\label{sec:backg}
The concept of trustworthiness management is evolving rapidly and has been widely employed in various disciplines \cite{hardin2002}\cite{schlenker1973effects}\cite{doi:10.1177/002224299405800302}\cite{mahmoodtrust} and applications (e.g., crowdsourcing, social recommendation)
\cite{7407618}\cite{7328306}\cite{6595455}. It is, therefore, important to distinguish the ideal optimal parameters for any IoT specific ecosystem.  
\subsection{Trust As A Concept}
\begin{figure}[!tb]
    \centering
    \includegraphics[width=\linewidth]{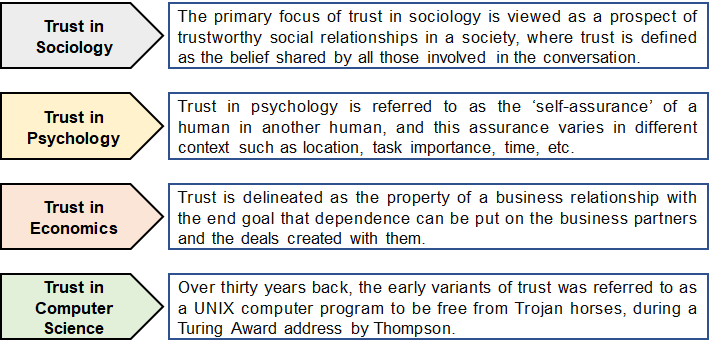}
    \caption{Trust definitions vis-à-vis different domains}
    \label{fig:trust_define}
    \vspace{-1em}
\end{figure}

Trust is a fundamental aspect of human life for building relationships with each other. With the rapid advancements (e.g., in terms of hardware and software) in science, the notion of trust is being integrated and utilized for different disciplines that require human behaviour analysis including but not limited to sociology, psychology, economics, and computer science \cite{10.1145/2501654.2501661}. The definition of trust varies with disciplines (Figure \ref{fig:trust_define}). In its basic form, trust is referred to as the belief of one human (trustor) on another human (trustee) \cite{10.1145/2501654.2501661}, and its notion relies on many facets, 
e.g., 
temporal factor, human propensity, and environmental conditions. 
A brief overview of trust in different domains is illustrated in this section.
\subsubsection{Trust in Sociology} Sociology studies human social relations, human societies, human-human interactions and the mechanisms that change and preserve these relations and societies \cite{doi:10.1177/1368431019831855}. The primary focus of trust in sociology is to ascertain trustworthy social relationships in a society, where trust is defined as the belief shared by all those involved in a conversation \cite{hardin2002}. Furthermore, the authors in \cite{deutsch1962cooperation}\cite{doi:10.1177/1368431003006002002} delineate trust as the mean of reducing the complexity in society, and it depends on the belief that a human places on the reactions of his/her counterpart. A different view of trust is provided in Seligman \cite{seligman2000problem}, wherein trust is described as the reliance and, in usable terms, it is a disposition with respect to the trustor to acknowledge reliance on a trustee.   
\subsubsection{Trust in Psychology} Psychology is a study of a human mind's characteristics, especially, in a specific context \cite{henriques2004psychology}. Trust in psychology is referred to as the self-assurance of a human in another human and this assurance varies in a different context, e.g., location, task importance, and time \cite{926617}. Likewise, various literature regards trust as a similar characteristic for both social science and psychology, however, the former accounts for trust in terms of societies, whereas, the 
latter
 deals with the same at an individual level \cite{hardin2002}\cite{schlenker1973effects}. Furthermore, Josang \etal \cite{JOSANG2007618} treats trust as the subjective behaviour via which an individual envisages that its counterpart accomplishes a given activity on which its assistance depends and termed it as $reliability \ trust$.
\subsubsection{Trust in Economics} Economics is also a part of social science that deals with the production and distribution of goods and services \cite{Backhouse2009}. Trust in economics is referred to as the reliability in business transactions, wherein one party has the belief in its counterpart's reliability and credibility \cite{doi:10.1177/002224299405800302}. In e-Commerce, it is possible to mitigate the transactions risk by incorporating trust dynamics via providing photos of the products, rating, and reviews when there is no direct interaction between the consumers and the products \cite{BA2002}\cite{10.1145/642611.642634}. Likewise, Kazuhiro \cite{Arai2007} delineates trust as the character of a business relationship with the end goal that the dependence can be put on the business partners and the deals created with them.    
\subsubsection{Trust in Computer Science} In computer science, the main strive is to (a) build a system that is secure, (b) fit for purpose, and (c) in face of any unexpected vulnerabilities, identify these vulnerabilities easily and recover efficiently \cite{ARTZ200758}. The current computer science systems are about data communication and processing that require secure and trustworthy management \cite{harper_2014}. In general, security is all about locks, gates, and fences, however, trust is regarded as when and where we need these enclosures and why they work for a particular environment \cite{harwood2012logic}. Moreover, the early variants of trust looks into various aspects of network and data security with one of the earliest by Thompson \cite{358210} delineating the trust as a UNIX computer program free from Trojan horses.  
\subsection{Characteristics of Trust}
Trust can be evaluated in numerous ways by considering the following characteristics \cite{9142365}:
\subsubsection{Subjective} Subjective trust, in terms of social perspective, is viewed as the evaluation of trust using the centrality of an object, wherein the trust is computed based on trustor's observation (i.e., direct trust) as well as the opinion (i.e., feedback or indirect trust) of the other objects.
\subsubsection{Objective} In contrast to subjective trust, an objective trust is evaluated by utilizing the feedback from all the objects in the network, wherein the trust information of each object is distributed and visible to everyone. Moreover, the accessibility of this information is possible via distributed hash tables and this information is maintained by pre-trusted social objects.  
\subsubsection{Local} It represents the trust based on an object-object relationship, wherein an object evaluates the trustworthiness of another object using local information such as its self-observation and past experience. 
\subsubsection{Global} In comparison to the local trust, the global trust is considered as the reputation of an object within the network, wherein the trust score of each object is computed by aggregating the local information of each of the other objects in the network.  
\subsubsection{Context-Specific} Trust of an object towards another object varies with context. A trust relation between the objects is usually dynamic and depends on multiple factors such as temporal factors, location, and energy status.  
\subsubsection{Asymmetric} Trust is an asymmetric property, i.e., if an object A trusts another object B, it does not guarantee that 
B also trusts 
A.    

\begin{figure*}[h]
    \centering
    \includegraphics[width=0.85\textwidth]{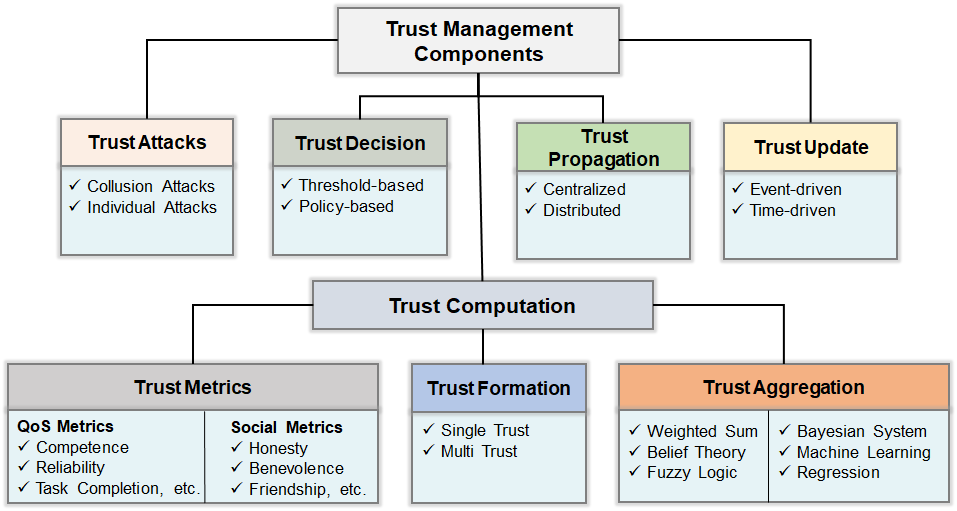}
     \caption{Components of trustworthiness management in SIoT}
    \label{fig:trust_mang}
   \vspace{-1em}
\end{figure*}

\subsection{Trust in SIoT}
As discussed, trust plays an important role in SIoT to make trustworthy decision independently without any human intervention. The paradigm of SIoT is more inclined towards social science and a commonly known characteristic of trust in this domain is the `‘\textit{confidence}’' or `‘\textit{belief}’' of an entity towards another entity \cite{app9010166}\cite{926617}. Thus, in SIoT, trust is widely acknowledged as \textit{the ``confidence'' of a trustor in a trustee to achieve an objective under a particular setting within a particular timespan} \cite{8254523}\cite{s17061346}. The concept of trust in SIoT is utilized in various applications, including but not limited to social Internet of Vehicles (SIoV) \cite{doi:10.1177/1550147719825820},  crowdsourcing \cite{7721739}, object recommendation \cite{8936977}, trustworthy service discovery \cite{10.1007/978-3-030-03596-9_9}, etc. 

In the SIoT paradigm, it is imperative to apprehend that a node (either a trustor or a trustee) can be an individual, a device, or an application. Subsequently, the assessment of trust can be a probability or a value, generally referred to as trust esteem. Furthermore, trust is neither the property of a trustor nor a trustee, 
in fact, 
it is the relationship between the two. The foremost objective of trust evaluation is to assess the action of the trustee (or the evaluation of the data it provides) as per the trustor’s prospect and trustee’s characteristic \cite{app9010166}\cite{article}. Thus, it is essential to consider the required parameter for trust quantification as the concept of trust is complex and can not be measured with a single parameter. Trust of a SIoT object can be seen as the degree of confidence or faith of various characteristics of an object, 
e.g., 
the object's ability, integrity, reliability, security, and dependability. Trust in SIoT can be seen as a reputation of an object in the SIoT network based on its direct and indirect understanding and previous transactions \cite{YAN2014120}. In general, the essential components to provide trustworthiness management in SIoT are portrayed in Figure \ref{fig:trust_mang} and are briefly discussed in Section~\ref{Sec:TMC}.  %

\renewcommand{\arraystretch}{1.25}
\newcolumntype{C}[1]{>{\centering\arraybackslash}p{#1}}
\begin{table*}[ht]
\footnotesize
\setlength{\arrayrulewidth}{.1em}
    \caption{Comparison of trust aggregations techniques}
    \label{tab:trust_aggregation}
    \begin{tabular}{ |>{\centering\arraybackslash}m{2.5cm} >{\centering\arraybackslash}m{5.25cm} >{\centering\arraybackslash}m{3.75cm} >{\centering\arraybackslash}m{4.5cm}|  }
     \hline
     \rowcolor{Gray}
      \small\textbf{Techniques} & \small\textbf{Description} & \small\textbf{Strengths} & \small\textbf{Weaknesses} \\ 
      \hline \hline
        Weighted Sum \small \textbf{\cite{stanimirovic2011linear}\cite{kim2006adaptive}\cite{marler2010weighted}} & \begin{tableitems}[nosep,after=\strut]
                        \item This aggregation technique refers to as an average weighted mean of each value, where each metric is assigned with the static or the dynamic weights to get the single score.
        \end{tableitems}  & 
        \begin{tableitems}[nosep,after=\strut]
            \item Low computations cost as it does not require any mathematical function. 
            \item it is a simple method of aggregating the values.
        \end{tableitems} &
        \begin{tableitems}[nosep,after=\strut]
            \item Infinite number of possibilities for determining the weights of each value different environments
            \item Inability to identify influence of each value on overall value
        \end{tableitems} \\[-2mm] \hline
        Belief Theory \small \textbf{\cite{Liu2008}\cite{Sen2002}\cite{beynon2000dempster}} & \begin{tableitems}[nosep,after=\strut]
                        \item It is also referred to as Dempster-Shafer theory or evidence theory. Belief theory combines multiple evidence and gives a degree of belief in range \{0,1\}, where 0 represents no support and 1 represents full support for the evidence.
        \end{tableitems}  & 
        \begin{tableitems}[nosep,after=\strut]
            \item This technique allows to combine data from different independent sources.
            \item Belief theory is appropriate for managing missing data and provides the better method of enumerating vagueness.  
        \end{tableitems} &
        \begin{tableitems}[nosep,after=\strut]
            \item In presence of malicious objects, the conflicting uncertainly in belief theory may disrupt the opinion of legitimate objects, and thus lead to unreliable decision making.  
        \end{tableitems}\\[-2mm] \hline
        Bayesian Inference \small \textbf{\cite{josang2002}\cite{weise1993bayesian}\cite{bernardo2009bayesian}} & \begin{tableitems}[nosep,after=\strut]
                        \item Bayesian inference is a popular technique for trust computational model where trust is treated as a random variable in the range of [0,1] following a beta distribution to designate the probability distribution of a data/node/interaction. The concept of Bayesian inference is based on Bayes’ theorem. \vspace{-1em}
        \end{tableitems}  & 
        \begin{tableitems}[nosep,after=\strut]
            \item It provides the solid theoretical framework to combine prior information with data. 
            \item The inferences in this technique are data dependent and are exact, without dependence on asymptotic estimation
        \end{tableitems} &
        \begin{tableitems}[nosep,after=\strut]
            \item Bayesian Inference requires more expertise to interpret prior distribution beliefs into a mathematically formulated prior distribution to avoid the misleading results.  
            \item Models with high number of parameters/metric often lead to high computational cost.  
        \end{tableitems}\\[-2mm] \hline
        Fuzzy Logic \small \textbf{\cite{6617083}\cite{zadeh1996fuzzy}\cite{carbo2003trust}} & \begin{tableitems}[nosep,after=\strut]
                    \item A fuzzy logic is a many-valued logic processing more than the customary two truth-values of truth and falsehood unlike Boolean logic. This concept of fuzzy logic is realistic for trust aggregation as the node cannot be characteristically labelled as completely trustworthy or completely untrustworthy.    
        \end{tableitems}  & 
        \begin{tableitems}[nosep,after=\strut]
                    \item Fuzzy logic resembles human reasoning that works well even in presence of ambiguous or vague input. 
                    \item This nature of fuzzy logic makes this approach suitable for complex nature of trust evaluation to make decision efficiently and effectively.  
        \end{tableitems} &
        \begin{tableitems}[nosep,after=\strut]
                \item Fuzzy logic system requires more testing and validation as one problem can have many potential solutions because of no any systemic approach and more human knowledge and expertise dependency. 
        \end{tableitems}\\[-2mm] \hline
        Machine Learning \small \textbf{\cite{8364607}\cite{sagar}\cite{10.1007/978-3-030-36808-1_56}} & \begin{tableitems}[nosep,after=\strut]
                        \item Machine learning-driven aggregation techniques are normally required two-step process for prediction; i) Unsupervised learning (Clustering) when the training data is not labelled, ii) Multi-class supervised learning (classification) to classify the interactions/nodes into different classes (i.e., trustworthy and/or untrustworthy). 
        \end{tableitems}  & 
        \begin{tableitems}[nosep,after=\strut]
                        \item This technique is more suitable if the number of trust metrics to compute the overall trust score increase when compared with other aggregation techniques.   
        \end{tableitems} &
        \begin{tableitems}[nosep,after=\strut]
                        \item It is computationally expensive to utilize machine learning-driven algorithm and it leads to high latency as the system needs to train the trust model after every transaction.   
        \end{tableitems}\\[-2mm] \hline
        Regression Analysis \small \textbf{\cite{Wang2014LogitTrustA}\cite{6397158}\cite{2017jay}} & \begin{tableitems}[nosep,after=\strut]
                        \item It is a statistical process that is used to approximate the relationship between the independent variable. Regression is divided into two categories, i.e., linear and multi-regression. Linear regression in terms of trust is defined as the value of trust depends on one independent variable/metric while in multiple regression, the trust is dependent on more than one independent variable. \vspace{-1em}  
        \end{tableitems}  & 
        \begin{tableitems}[nosep,after=\strut]
                        \item Multi-regression analysis has the capability to determine the impact of each trust metric while aggregating the multiple metrics, and is able to identify the outlier more efficiently. 
        \end{tableitems} &
        \begin{tableitems}[nosep,after=\strut]
                        \item Regression may leads to the uncertain results when the dataset used for analysis is insignificant.
                        \item Linear regression usually oversimplify the problem, and thus, it is not recommended for real-world complex problem.    
        \end{tableitems}\\
      \hline 
    \end{tabular}
 \vspace{-3mm}
\end{table*}

\section{Trust Management Components}
\label{sec:tmc}
This section presents the essential components that are to be considered for trust management process in SIoT. 
\label{Sec:TMC}
    \subsection{Trust Computation}
        \subsubsection{Trust Metrics} Trust metrics refer to the features that are chosen and combined for trust purposes. These features can be chosen in terms of a node's social trust metrics and/or quality of service (QoS) trust metrics.
            \begin{itemize}
                \item \textit{Social Metrics:} The social trust metrics represent the social behaviour of nodes in terms of the social relationship between the owners of IoT devices and is measured using integrity, benevolence, honesty, friendship, community-of-interest, and unselfishness \cite{Nitti2014}\cite{7097037}\cite{8737491}.
                \item \textit{QoS Metrics:} It represents the confidence that a node is able to offer the QoS and is measured in terms of reliability, competence, data delivery ratio, throughput, and task completion \cite{8708985}\cite{8070939}\cite{7289151}.
            \end{itemize}
        \subsubsection{Trust Formation} Trust formation forms the trust either based on a single aspect, i.e., in terms of positive or negative QoS or multiple aspects, i.e., trust models that include both QoS and social trust metrics.  
            \begin{itemize}
                \item \textit{Single Trust:} Single trust represents the fact that only single trust metric (e.g. quality of service metric) is used to ascertain the overall trust \cite{6940301}\cite{doi:10.1155/2015/859731}\cite{BENSAIED2013351}.
                \item \textit{Multi Trust:} It employs the notion of trust as a multi-dimensional concept. For instance, combining multitude of factors like both social and QoS metrics to form a single trust score \cite{article}\cite{7097037}\cite{8737491}.  
            \end{itemize}
        \subsubsection{Trust Aggregation} It consists of techniques that aggregate trust observation to obtain a single trust score. Many aggregation techniques have been investigated in the research literature \cite{CHAHAL202013}, including but not limited to, the one based on weighted sum \cite{7097037}\cite{8653859}, belief theory \cite{beynon2000dempster}\cite{10.1145/544741.544809}, Bayesian system \cite{6940301}\cite{6513398}, fuzzy logic \cite{6617083}\cite{8737491}, regression analysis \cite{Wang2014LogitTrustA}\cite{6397158}, and machine learning \cite{sagar}\cite{9322540}. Trust aggregation is an important step of any trust computation model, and therefore, it is pertinent to discuss the trust aggregation techniques in a comparative manner. Table \ref{tab:trust_aggregation} illustrates each of these aggregation techniques along with the strengths and weaknesses of the same. 
        \begin{itemize}
            \item \textit{Weighted Sum}: This technique is the simplest and one of the commonly used aggregation method. The technique refers to   as an average weighted mean of each metric/value, where each metric is assigned with a weight to get the single score. 
            Let $M=\{m_1, m_2, m_3,...,m_n\}$ represents the $n$ trust metrics and $W=\{w_1, w_2, w_3,...,w_n\}$ represent the weights of each $n$ trust metrics \cite{7097037}\cite{8653859}, the weighted sum aggregation $(WS_A)$ is computed as:
            \begin{equation}
                WS_A = \sum\limits_{i=1}^n W_i * M_i
                \label{eq:weightedsum}
            \end{equation}
            Here the weights can be either \emph{static}, i.e., the weights remain the same for each metric or \emph{dynamic}, i.e., the weights can change over time.  
            
            \item \textit{Belief Theory}: It is also referred to as Dempster-Shafer Theory (DST) or evidence theory. Belief theory combines multiple evidence and gives a degree of belief in range \{0,1\}, where 0 represents no support and 1 represents full support for the evidence. DST provides an uncertainty interval in terms of belief ($bel$) and plausibility ($pla$) instead of a traditional probability \cite{beynon2000dempster}\cite{10.1145/544741.544809}. The belief of a node $\mathcal{N_j}$ in view of node $\mathcal{N_i}$ with respect to an event $\mathfrak{a}$ is computed as: 
            \begin{equation}
                bel(\mathcal{N_i}) = \sum\limits_{a_j \subseteq a} m_{\mathcal{N_i}}(\mathfrak{a_j})
                \label{eq:belief}
            \end{equation}
            
            Here $\mathfrak{a_j}$ represents all the basic events of $\mathfrak{a}$, and $m_{\mathcal{N_i}}(\mathfrak{a_j})$ highlights all the events in view of $\mathcal{N_i}$. Therefore, we can conclude that the belief of a node for an event $\mathfrak{a}$ is    $bel(\mathcal{N_j}) = m_{\mathcal{N_i}}(\mathfrak{a})$. Subsequently the plausibility is $pla(\mathcal{N_i}) = 1- bel(\mathcal{N_i})$. 
            
            \item \textit{Bayesian System}: The concept of Bayesian system is based on Bayes' theorem, i.e., the prior probability, posterior probability about the data/node/interaction, and the likelihood function. The trust in Bayesian system is treated as the random variable and is stated as follows \cite{6940301}\cite{6513398}:
            \begin{equation}
                p(\mathcal{A}|\mathcal{B}) = \frac{p(\mathcal{B}|\mathcal{A})p(\mathcal{A})}{p(\mathcal{B})} 
                \label{eq:bayesian}
            \end{equation}
            Here $p(\mathcal{A}|\mathcal{B})$ is the posterior probability of $\mathcal{A}$ given $\mathcal{B}$ is true, $p(\mathcal{B}|\mathcal{A})$ is the likelihood of $\mathcal{B}$ given $\mathcal{A}$ is true, $p(\mathcal{B})$ is the probability of $\mathcal{A}$ happening, and  $p(\mathcal{A})$ is the prior probability of $\mathcal{A}$.  
            
            \item \textit{Fuzzy Logic}: In contrast to the Boolean logic which takes precise input in the form of $0$ or $1$, fuzzy logic provides a more realistic understanding similar to human reasoning. Accordingly, fuzzy logic can address the uncertainly and fuzziness in notion of trust \cite{6617083}\cite{8737491}. In general, a fuzzy aggregation technique can be divided into following four phases:
            i) \textit{Fuzzy Controller --} to transform the real values into fuzzy sets, 
            ii) \textit{Fuzzy Logic Rules --} to design the fuzzy logic rules via employing fuzzy intersection, fuzzy union, etc.,  
            iii) \textit{Membership Function (Mapping Function) --} to transform the fuzzy input sets into fuzzy output sets, and 
            iv) \textit{Defuzzy Controller --} to convert the fuzzy output sets into the real values.  
            
            \item \textit{Regression Analysis}: This statistical process utilizes the slope of the lines to aggregate different independent variables. Regression is divided into two types: i) \textit{Linear regression}, to make the prediction about one dependent variable based on the information available for one independent variable, and ii) \textit{multi regression}, to predict the output of a dependent variable based on the information available from many independent variables \cite{Wang2014LogitTrustA}\cite{6397158}. Mathematically, the linear regression can be seen as
             %
            \begin{equation}
                \mathcal{Y} = m_0 + m_1 \mathcal{X}
                \label{eq:l_regression}
            \end{equation}
            and multi regression is computed as follows:
            \begin{equation}
                \mathcal{Y} = m_0 + m_1 \mathcal{X}_1 + m_2 \mathcal{X}_2 + ... + m_n \mathcal{X}_n
                \label{eq:m_regression}
            \end{equation}
            Here $m_0$ represents the y-intercept of the line, $m_1, m_2,...,m_n$ are the slops of the lines, $\mathcal{Y}$ is the dependent variable (i.e., aggregated score), and $\mathcal{X}_1, \mathcal{X}_2, ... ,\mathcal{X}_n $ are the independent variables (i.e., trust metrics for trust computation).
            
            \item \textit{Machine Learning}: Machine learning-based aggregation techniques utilize the clustering (i.e., unsupervised algorithms) and classification (i.e., supervised algorithms), and are data dependent. If the data is not labelled then aggregation requires two steps: i) unsupervised algorithms (e.g., k-mean clustering, agglomerative clustering, and spectral clustering) to label the data, and ii) supervised algorithms (e.g., support vector machine, logistic regression, and random forest) to classify the nodes/objects as either trustworthy or untrustworthy \cite{sagar}\cite{9322540}.     
        \end{itemize}

     \subsection{Trust Propagation} Trust propagates facilitates in understanding that how the trust propagates in the network and is generally categorized in following 
     three 
     broad schemes:  
        \begin{itemize}
            \item \textit{Centralized:} Centralized schemes rely on a centralized entity which is primarily responsible for (a) gathering trust-related information for the purpose of trust computation and (b) propagating the same in the network \cite{Nitti2014}\cite{BENSAIED2013351}. However, centralized controlled frameworks are vulnerable to a single point of failure, wherein the entire network can collapse.
            
            \item \textit{Distributed:} In distributed schemes, objects are responsible for both trust computation and propagation within the network without any centralized authority \cite{6940301}\cite{doi:10.1002/dac.2930}. This scheme although provides a solution to the single point of failure, nevertheless, has inherent challenges, i.e., honest trust computation, managing computational capabilities and the unbiased trust propagation in the entire network. 
            
            \item \textit{Hybrid:} Hybrid schemes are generally used to overcome the challenges posed by both centralized and distributed schemes. Furthermore, hybrid schemes divide the propagation in two common categories, i.e., \textit{locally distributed and globally centralized} and \textit{locally centralized and globally distributed} \cite{Nitti2014}\cite{8468022}. 
        \end{itemize}    
        
    \subsection{Trust Update} At the end of a transaction or at any specified time interval, trust score of a trustee is updated based on it performance. Thus, the update can take place in 
    three 
    ways: 
        \begin{itemize}
            \item \textit{Event-driven:} In this approach, trust is updated after each transaction or once an event has occurred \cite{6940301}\cite{BENSAIED2013351}. Nevertheless, this type of update increases the traffic overhead in networks with more frequent transactions.  
            \item \textit{Time-driven:} In time-driven approach, trust is collected and updated periodically after a given interval of time \cite{doi:10.1002/dac.2930}\cite{7383635}. Although this approach overcomes the problem of event-driven approaches, nevertheless, selecting an appropriate time interval remains a challenge.   
            \item \textit{Hybrid:} A number of studies consider both the event-driven and time-driven approaches for trust update where a trust is updated periodically and/or in case of an event (after an interaction) \cite{7289151}.   
        \end{itemize}

\begin{figure}[t]
    \centering
    \includegraphics[width=\linewidth]{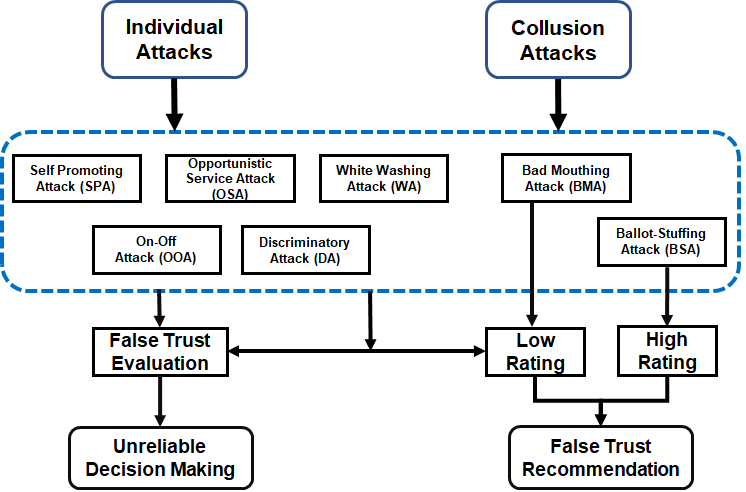}
    \caption{Trust related attacks}
    \label{fig:trust_attack}
    \vspace{-1em}
\end{figure}

    \subsection{Trust Related Attacks}
        A node can act maliciously with an intent to break the basic functionality of the network and services. There are two types of trust-related attacks as depicted in Figure \ref{fig:trust_attack}. These attacks are categorized as individual attacks and collusion-based attacks \cite{MS201932}\cite{electronics10182223}.
        \begin{itemize}
        \item \textit{Individual Attacks:}
        Individual attacks refer to the attack launched by an individual object. Some of the common form of individual attacks are briefly discussed as follows:
        \begin{itemize}
            \item \textit{Self Promoting Attacks (SPA):} In this type of attacks, a node can promote its significance by providing regular good recommendation for itself so as to be selected as a service provider, and once the node is selected as a service provider, it acts maliciously \cite{CHAHAL202013}.
            \item \textit{Whitewashing Attacks (WA):} In a white washing attack, a node can exit and re-join the network or an application to recover its reputation and to wash-away its own bad reputation.
            \item \textit{Discriminatory Attacks (DA):} In DA, a node explicitly attacks other nodes that do not have various common friends by virtue of human intuition or affinity towards friends in SIoT structures. This attack is sometimes referred to as selective behaviour attack where a node performs well for a particular service/node and ineffectually for some other services/nodes \cite{10.1007/978-3-030-34986-8_28}.
            \item \textit{Opportunistic Service Attacks (OSA):} An object can intelligently offer a great service to improve its reputation when its reputation falls in light of offering a bad service. With a high reputation, an object can collude with different objects to perform collusion-based attacks \cite{9264256}.   
            \item \textit{On-Off attacks (OOA):} OOA is similar to OSA, however, in these sorts of attacks, an object provides good and bad services on and off (randomly) to avoid being labelled as a low reputed node, thereby increasing its chance of being selected as a service provider \cite{CHAHAL202013}. 
        \end{itemize}
        
        \item \textit{Collusion-based Attacks:} Collusion attacks represent the attack launched by group of objects to either provide the high rating or low rating to a particular object. Following are some of the collusion attacks: 
        \begin{itemize}
            \item \textit{Bad Mouthing Attacks (BMA):} In BMA, a node can deteriorate the reputation of a trustworthy node within the network by providing bad recommendations to diminish its chance of being chosen as a service provider \cite{9305298}.
            \item \textit{Ballot Stuffing Attacks (BSA):} These attacks are used for boosting the reputation of bad nodes within the network by providing good recommendations for them so that the bad node can be selected as a service provider \cite{9305298}. 
        \end{itemize}
        \end{itemize}

    \subsection{Trust Decision} After computing the trust score of a trustee, the main purpose of devising a trust management system is to identify whether a node is trustworthy or untrustworthy by means of using any of the following two techniques:
        \begin{itemize}
            \item \textit{Threshold-based Decision:} In threshold-based decision techniques, decision is taken on the basis of either a rank-based function or a threshold value. \cite{Chakrabarti2007}\cite{8002573}. Moreover, the threshold values can be adaptive so as to facilitate dynamic environments, whereas static values are specifically employed for a particular application or service.  
            \item \textit{Context-based Decision:} This technique forms the policies that are used to identify and decide whether an object is classified as malicious or not by using the contextual information in terms of location, temporal factor, energy status, etc \cite{Chakrabarti2007}\cite{7959573}.
        \end{itemize}   
%
%
%

\section{Trust Management Schemes: Discussion and Analysis}
\label{sec:TMS}
As of late, there has been an increased interest of the research community to provide an insight on trustworthiness management systems for SIoT. Therefore, this survey categorizes these systems into four broad categories, namely \textit{recommendation-based}, \textit{reputation-based}, \textit{prediction-based}, and \textit{policy-based schemes}. This section compares the trust management systems that have already been described in the literature, first discussing the pros and cons of these schemes as well as whether or not they are context-aware. Subsequently, these methods are classified based on the trust metrics employed for trust quantification and the trust-related attacks that they manage.

\subsection{Discussion on Trust Management Schemes}
This section presents the detailed discussion on the four categorized trust management schemes.

\subsubsection{Recommendation-based Trust Scheme (RecTS)}
Over the years, a number of recommendation-based trust management systems have been employed, wherein recommendations as a trust metric is exploited in a bid to evaluate the trustworthiness of the nodes in a SIoT network \cite{articlemohamadi}. The trust decision in these approaches is based on both the direct observations as well as the recommendations from the neighbouring nodes to make a more precise decision (even if there is no current direct observations or previous direct observations \cite{sfar2018roadmap}. As of late, a number of trust management systems employing recommendation as a trust metrics are proposed \cite{Nitti2014}\cite{10.1007/978-3-030-03596-9_9}\cite{7097037}\cite{8737491}\cite{doi:10.1002/dac.2930}\cite{9211717} in the research literature. 

Nitti \etal \cite{Nitti2014} present the trustworthiness management model for SIoT, employing both the subjective and objective properties of an object. The subjective model is derived by considering the social point of view of an object in terms of its centrality, its own direct experience, and the opinions of neighbouring friends. The objective trustworthiness is employed as a notion of the peer-to-peer network, wherein the information of each object is stored in a distributed hash table and is visible to every object in the network. The computation of objective trustworthiness involves centrality, long and short-term opinions from all the objects in the network. Finally, the static weighted sum aggregation is employed to compute the single trust score. Similarly, the adaptive trust model, suitable for the dynamic changing environment in SIoT is introduced by Chen \etal \cite{7097037} to isolate the misbehaving nodes performing trust-related attacks. The model considers honesty, cooperativeness, and community-of-interest as the trust metrics, and the recommendation is considered as the direct trust of neighbouring objects. To defend against the good and bad-mouthing from any recommender, a dynamic parameter is considered to control the impact of recommendations for computing the trustworthiness of an object. The performance evaluation of the model is carried out in terms of convergence, accuracy, and resiliency.

Furthermore, Xia \etal \cite{8737491} delineate a context-aware trustworthiness inference framework by employing two trust metrics, \textit{similarity trust} and \textit{familiarity trust}. 
The familiarity trust considers the kernel-based nonlinear multivariate grey prediction model to compute the direct trust, and recommendations as indirect trust. The similarity trust is computed by employing centrality, and community-interest metrics. To synthesize both the trust metric, a fuzzy logic based aggregation technique is proposed to get the single trust score. The validity of the model is considered in terms of its resistance to numerous trust-related attacks. Khani \etal \cite{10.1007/978-3-030-03596-9_9} present a mutual context-aware trust evaluation model, wherein three social trust metrics and two QoS metrics are considered to evaluate the trustworthiness of an object. The three social metrics are social similarity in terms of friendship, community-of-interest, and relationships, and QoS metrics are expected and advertised QoS.  For context-awareness, the status of a device (energy and computation capability), environment (location), and task type are integrated for trust metrics computation. Finally, the static weighted sum aggregation approach is employed to segregate these independent trust metrics. 

Most recently, Wei \etal \cite{9211717} propose a context-aware socio-cognitive-based trust model for service delegation in service-oriented SIoT. The model is based on two characteristics, \textit{competence quantification} and \textit{willingness quantification}. The competence is quantified by utilizing the degree of importance (DoI) and degree of social relationships (DoSR), and willingness quantification integrates the degree of contribution (DoC) and also the DoSR. The DoI quantifies service providers' competency in terms of computational power, storage, and communication capabilities, the DoC ensures the willingness of the service provider, and the DoSR is employed as the weighting parameters for both competence and willingness. The final trust score is computed by aggregating both the trust parameters by using the weighted sum technique.   

\renewcommand\arraystretch{1.25}
\begin{table*}[t]
	\footnotesize
\setlength{\arrayrulewidth}{.075em}
  \begin{center}
    \caption{Trust computation techniques}
    \label{tab:recommendation}
    \setlength{\leftmargini}{0.3cm}
   \begin{tabular}{|>{\centering\arraybackslash}m{1.75cm}| >{\centering\arraybackslash}m{2.9cm}| >{\centering\arraybackslash}m{5cm}| >{\centering\arraybackslash}m{5cm}| >{\centering\arraybackslash}m{1.25cm}|}
     \hline 
     \multicolumn{5}{c}{\text{\textbf{\normalsize Recommendation-based Techniques}}} \\ 
     \hline
     \rowcolor{Gray}
       \small\textbf{Ref.} & \small\textbf{Trust metrics} & \small \textbf{Strengths} & \small \textbf{Limitations} & \textbf{Context-awareness} \\ 
      \hline \hline
      \small Nitti \etal$\textbf{\cite{Nitti2014}}$ & \begin{tableitems}[nosep,after=\strut]
                        \item Centrality
                         \item Credibility
                        \item Feedback/opinion
        \end{tableitems} 
        & \begin{tableitems}[nosep,after=\strut]
                        \item The envisaged trust model encompasses both the subjective and the objective properties of a node within the network.
                        \item Efficient in terms of successfully identifying malicious nodes in the network even in presence of a high percentage of malicious nodes. \vspace{-1em}
        \end{tableitems} 
        & \begin{tableitems}[nosep,after=\strut]
                        \item Feedback increases the network traffic overhead due to transmission.
                         \item Model responds slowly to dynamic changes in the environment due to pre-defined trust parameters. 
        \end{tableitems}  & No \\ \hline
                \small Khani \etal$\textbf{\cite{10.1007/978-3-030-03596-9_9}}$ & \begin{tableitems}[nosep,after=\strut]
                        \item QoS metric 
                        \item Friendship
                        \item Community-of-interest
                        \item Feedback as recommendations 
        \end{tableitems} 
        & \begin{tableitems}[nosep,after=\strut]
                        \item This framework considers mutual context-awareness while computing the trust and outperform many existing trust related models.
                        \item The model performs well against many trust related attacks, including but not limited to BMA, BSA, SPA, and OOA. \vspace{-2em}
        \end{tableitems} 
        & \begin{tableitems}[nosep,after=\strut]
                        \item Fine tuning of trust parameters in lieu of dynamically changing environments has not been taken into consideration. 
        \end{tableitems}  & Yes \\ \hline
       \small Chen \etal$\textbf{\cite{7097037}}$ & \begin{tableitems}[nosep,after=\strut]
                        \item Honesty 
                        \item Cooperativeness
                        \item Community-of-interest
                        \item Recommendations
        \end{tableitems} 
        & \begin{tableitems}[nosep,after=\strut]
                        \item The model is suitable for the dynamically changing environmental conditions.
                        \item The proposed model fine tunes the trust parameters primarily depending on dynamically changing environments. \vspace{-1em}
        \end{tableitems} 
        & \begin{tableitems}[nosep,after=\strut]
                        \item Defense mechanism against some key attacks e.g., on-off and intelligent behaviour attack, have not been considered.
                        \item The selection and fine tuning of some of the important parameters is still an issue in the proposed model. \vspace{-1em}
        \end{tableitems}  & No \\ \hline
        \small\ Xia \etal$\textbf{\cite{8737491}}$ & \begin{tableitems}[nosep,after=\strut]
                        \item Centrality
                        \item Recommendations
                        \item Community-of-interest
        \end{tableitems} 
        & \begin{tableitems}[nosep,after=\strut]
                        \item The model consider context-awareness and outperform two well known trust models relying on weighted sum approach.
                        \item The proposed model performs well in presence of grey-hole attacks and bad-mouthing attacks. \vspace{-1em}
        \end{tableitems} 
        & \begin{tableitems}[nosep,after=\strut]
                        \item Integration of context information and discussion on weights for different attributes is missing. 
                        \item Defense mechanisms against the on-off attacks and intelligent behaviour attacks have not been considered. \vspace{-1em}
        \end{tableitems}  & Yes \\ \hline
        \small Wei \etal$\textbf{\cite{9211717}}$ & \begin{tableitems}[nosep,after=\strut]
                        \item Direct trust
                        \item Experience
        \end{tableitems} 
        & \begin{tableitems}[nosep,after=\strut]
                        \item A context-aware socio-cognitive based trust model for service delegation is proposed in this paper where the trust quantification process involve two adaptable and dynamic trust characteristic in the form of degree of contribution and degree of importance. 
                        \item The proposed model has high success rate of tasks in presence of many trust-related attacks such as OSA, BMA and BSA. \vspace{-1em}
        \end{tableitems} 
        & \begin{tableitems}[nosep,after=\strut]
                        \item With increase in number of properties included for trust computation, the convergence time of the model also increases rapidly.
                        \item The proposed model integrates many trust attributes to compute the trust score, however, the weighted sum aggregation lacks in identifying the appropriate weight for each attribute. \vspace{-1em}
        \end{tableitems}  & Yes \\ 
      \hline 
    \end{tabular}
  \end{center}
  \vspace{-3mm}
\end{table*}

\begin{table*}[ht]
\footnotesize
\setlength{\arrayrulewidth}{.075em}
  \begin{center}
    \caption{Trust computation techniques}
    \label{tab:reputation}
    \setlength{\leftmargini}{0.3cm}
   \begin{tabular}{|>{\centering\arraybackslash}m{1.75cm}| >{\centering\arraybackslash}m{2.9cm}| >{\centering\arraybackslash}m{5cm}| >{\centering\arraybackslash}m{5cm}| >{\centering\arraybackslash}m{1.25cm}|}
     \hline 
             \multicolumn{5}{c}{\text{\textbf{\normalsize Reputation-based Techniques}}} \\ 
             \hline
             \rowcolor{Gray}
            \small \textbf{Ref.} & \small \textbf{Trust metrics} & \small \textbf{Strengths} & \small \textbf{Limitations} & \textbf{Context-awareness} \\ 
      \hline \hline
        \small Truong \etal$\textbf{\cite{8254523}}$ & \begin{tableitems}[nosep,after=\strut]
                        \item Reputation 
                         \item Experience
        \end{tableitems} 
        & \begin{tableitems}[nosep,after=\strut]
                        \item The proposed model presented two trust metrics, i.e., experience and reputation. The experience is discussed in terms of cooperative, uncooperative and neutral interaction, whereas the reputation involve the positive and negative reputation. 
                        \item The performance evaluation is analysed via change in the experience over time in term of weak and strong tie and the convergence of the algorithm. \vspace{-1em} 
        \end{tableitems} 
        & \begin{tableitems}[nosep,after=\strut]
                        \item Analysis of individual trust metrics is present in this paper, however, the combined effect of these metrics to compute the trust of an object is not present.  
                        \item The performance evaluation in terms of trust computation and trust-related attacks are not present. 
        \end{tableitems}  & No \\ \hline
        \small Xiao \etal$\textbf{\cite{7289151}}$ & \begin{tableitems}[nosep,after=\strut]
                        \item Reputation 
                         \item Credit
        \end{tableitems} 
        & \begin{tableitems}[nosep,after=\strut]
                        \item This model is computationally optimal as it only requires two metrics (i.e., credit and reputation) to categorize the nodes as trustworthy or untrustworthy.
                        \item All the computations are done by reputation server so the nodes are not responsible for any type of computation. \vspace{-1.25em} 
        \end{tableitems} 
        & \begin{tableitems}[nosep,after=\strut]
                        \item All the aspects of trust computation such as trust composition, direct observations, recommendations, etc., are not considered. 
                        \item Due to centralized computation system, the scalability is the major concern owing to the scalable nature of SIoT. \vspace{-1.25em}
        \end{tableitems}  & No\\ \hline
        \small Chen \etal$\textbf{\cite{doi:10.1002/dac.2930}}$ & \begin{tableitems}[nosep,after=\strut]
                        \item Reputation
                        \item Social relationships
                        \item Energy status of a device
        \end{tableitems} 
        & \begin{tableitems}[nosep,after=\strut]
                        \item Model considers energy-aware technique to balance the workload in the network, and encompasses energy of node, past performance (i.e., reputation) and social relationships to ascertain the trust score.
                        \item For dynamic performance enhancement, timeliness of each transaction/interaction is also considered as an evaluation metric. \vspace{-2em}
        \end{tableitems} 
        & \begin{tableitems}[nosep,after=\strut]
                        \item Model does not consider any defense mechanism against trust related attacks.
                        \item The comparison of energy consumption is analyzed in this model, however, the effect of energy consumption with increase in number of nodes is an important factor that needs to be considered. \vspace{-2em} 
        \end{tableitems}  & No\\ \hline
        \small Azad \etal$\textbf{\cite{8949450}}$ & \begin{tableitems}[nosep,after=\strut]
                        \item Reputation
                        \item Experience
        \end{tableitems} 
        & \begin{tableitems}[nosep,after=\strut]
                        \item A decentralized trust management model is proposed that not only provides the trustworthiness of object in SIoT but also integrates the homomorphic encryption to protect the privacy of the objects.
                        \item The model works in self-enforcing manner, thus, there is no need of any trusted third party. \vspace{-1em}
        \end{tableitems} 
        & \begin{tableitems}[nosep,after=\strut]
                        \item Performance evaluation of the model has not been carried out in the presence of trust related attacks. 
        \end{tableitems}  & No\\ \hline
        \small Truong \etal$\textbf{\cite{nguyen2016}}$ & \begin{tableitems}[nosep,after=\strut]
                        \item Reputation
                        \item Knowledge
                        \item Experience
        \end{tableitems} 
        & \begin{tableitems}[nosep,after=\strut]
                        \item A fuzzy-based trust model is presented by taking into consideration the knowledge, reputation and experience as a trust metrics.
                        \item To analyse the applicability of the model, a trust car sharing use case is considered by employing the three metrics, i.e, reliability, pricing and quality. \vspace{-1em}
        \end{tableitems} 
        & \begin{tableitems}[nosep,after=\strut]
                        \item The evaluation of model in terms of detecting the misbehaving objects, accuracy of model, and convergence is not present.
                        \item Performance in terms of trust-related attacks is also not considered. 
        \end{tableitems}  & No\\
      \hline
    \end{tabular}
  \end{center}
  \vspace{-3mm}
\end{table*}

Conclusively, the recommendation-based trust model has numerous advantages as shown in Table \ref{tab:recommendation} including but not limited to evaluation of trust when there is no previous communication or the direct observation among the nodes is present, to include the importance and influence of the credible nodes in the network before relying on the information provided by them, etc. Nonetheless, quantifying the credibility of a node in a dynamic environment and the defence mechanism against recommendation-based attacks (e.g., 
BSA and BMA) is still a major challenge.   

\begin{table*}[h]
\footnotesize
\setlength{\arrayrulewidth}{.075em}
  \begin{center}
    \caption{Trust computation techniques}
    \label{tab:prediction}
    \setlength{\leftmargini}{0.3cm}
   \begin{tabular}{|>{\centering\arraybackslash}m{1.75cm}| >{\centering\arraybackslash}m{2.9cm}| >{\centering\arraybackslash}m{5cm}| >{\centering\arraybackslash}m{5cm}| >{\centering\arraybackslash}m{1.25cm}|}
     \hline 
             \multicolumn{5}{c}{\text{\textbf{\normalsize Prediction-based Techniques}}} \\ 
             \hline
             \rowcolor{Gray}
            \small \textbf{Ref.} & \small \textbf{Trust metrics} & \small \textbf{Strengths} & \small \textbf{Limitations} & \textbf{Context-awareness} \\ 
      \hline \hline
        \small Jayasinghe \etal $\textbf{\cite{8364607}}$ & 
        \begin{tableitems}[nosep,after=\strut]
                        \item Community-of-interest
                        \item Friendship similarity
                        \item Centrality
                        \item Cooperativeness \vspace{-1em}
        \end{tableitems} 
        & \begin{tableitems}[nosep,after=\strut]
                        \item A data-centric trust evaluation model is proposed which utilizes the machine learning-based trust aggregation to ascertain the single trust metrics instead of a weighted sum aggregation. \vspace{-1em}
        \end{tableitems} 
        & \begin{tableitems}[nosep,after=\strut]
                        \item Performance evaluation of the proposed model has not been carried out vis-\'a-vis various trust-based attacks.
                        \item The model suffers with scalability and reliability issue with an increase in the number of nodes.  \vspace{-1em}
        \end{tableitems}  & No \\ \hline
                 \small Marche \etal $\textbf{\cite{9305298}}$ & \begin{tableitems}[nosep,after=\strut]
                        \item Goodness score
                        \item Usefulness score
                        \item Preseverance score \vspace{-1em}
        \end{tableitems} 
        & \begin{tableitems}[nosep,after=\strut]
                        \item An incremental SVM-based trust-related attack detection model is proposed by employing three score as the indicator of trust: goodness, usefulness, and perseverance score. 
                        \item All the trust-related attacks are considered to analyse the performance of the model.\vspace{-2.25em}
        \end{tableitems} 
        & \begin{tableitems}[nosep,after=\strut]
                        \item The proposed model performs better in a network with mix of different attacks, nevertheless, performance degrades when individual attacks are considered.  
                        \item The discussion on various simulation parameters considered for performance evaluation is not known. \vspace{-1.25em}
        \end{tableitems}  & No \\ \hline
        \small Aalibagi \etal$\textbf{\cite{9328463}}$ & \begin{tableitems}[nosep,after=\strut]
                        \item Similarity
                        \item Centrality
        \end{tableitems} 
        & \begin{tableitems}[nosep,after=\strut]
                        \item A matrix factorization-based trust model is proposed that utilizes bipartite graph, the hellinger distance and the matrix factorization to identify trustworthy nodes.
                        \item  The model mitigates the cold start problem and performs well under malicious objects, specifically under OSA. \vspace{-1em}
        \end{tableitems} 
        & \begin{tableitems}[nosep,after=\strut]
                        \item The discussion on suitability of bipartite graph over other type of graphs (e.g., hyper graph) is not mentioned.
                        \item Evaluation of trust model under various other trust-related attacks is not known. \vspace{-1em}
        \end{tableitems}  & No \\ \hline
        \small Sagar \etal$\textbf{\cite{sagar1}}$ & \begin{tableitems}[nosep,after=\strut]
                        \item Community-of-interest
                        \item Cooperativeness
                        \item Friendship similarity
                        \item Co-work similarity
        \end{tableitems} 
        & \begin{tableitems}[nosep,after=\strut]
                        \item A machine learning-driven trust evaluation model has been proposed so as to observe the behaviour of nodes over a period of time.
                        \item The model has also analyzed the impact of each trust metric on the overall trust score. \vspace{-1em} 
            \end{tableitems} 
        & \begin{tableitems}[nosep,after=\strut]
                        \item Computationally expensive and results in high latency in highly dynamic environment owing to training the machine learning model more frequently.
                        \item Performance evaluation of the proposed model has not been carried out vis-\'a-vis various trust-based attacks. \vspace{-1em}
        \end{tableitems}  & No \\ \hline
        \small Abderrahim \etal$\textbf{\cite{7986378}}$ 
        & \begin{tableitems}[nosep,after=\strut]
                        \item Sociability
                        \item Direct observation
                        \item Recommendations
        \end{tableitems} 
        & \begin{tableitems}[nosep,after=\strut]
                        \item A scalable and adaptive community-based trust model is proposed to detect trust attacks especially the On-Off attack. 
                        \item The Kalman filter technique is used to estimate the behaviour of a node.  \vspace{-1em}
        \end{tableitems} 
        & \begin{tableitems}[nosep,after=\strut]
                        \item The validity of the model is evaluated on the basis of On-Off attack, however, it is important to investigate the behaviour on other trust related attacks. \vspace{-1em}
        \end{tableitems}  & No \\
      \hline
    \end{tabular}
  \end{center}
  \vspace{-3mm}
\end{table*}

\subsubsection{Reputation-based Trust Scheme (RepTS)}
The concept of reputation has been widely used for the dynamic IoT environment where devices/nodes are susceptible to risks owing to incomplete and manipulated information. The reputation of a node can be seen as a behaviour expectation towards other nodes based on experience and the collected referral information \cite{DEMEO201858}. Recently, reputation-based systems have been employed in many fields of computer science including but not limited to distributed networks, peer-to-peer networks, IoT environment
where security, privacy, and trust are the critical issues \cite{9043531}. Many reputation-based trust models \cite{8254523}\cite{7289151}\cite{doi:10.1002/dac.2930}\cite{8949450}\cite{nguyen2016}\cite{JAFARIAN2020107254}\cite{8885376}\cite{7816942} are employed to enhance the trustworthiness evaluation of a node and to detect the misbehaving nodes in the SIoT network.

Truong \etal \cite{8254523} present a trust model, referred to as REK wherein the experience and reputation are employed as an indicator of trust of an object. The computation of experience involves three factors: 1) intensity of interactions, 2) values of interaction in terms of cooperative, uncooperative and neutral, and 3) current state of relationships. Subsequently, the trend of experience is analyzed via development of experience (due to cooperative interaction), loss of experience (due to uncooperative interactions) and decay of experience (due to no or neutral interactions). The repudiation perspective of trust involves the concept of \emph{Google\ PageRank} algorithm wherein both positive and negative reputation are considered to compute the overall reputation of an object. Finally, the model is evaluated in terms of its convergence with minimum iterations. Xiao \etal in \cite{7289151} propose an optimal credit and reputation-based trust model for SIoT wherein two parameters credit (referred to as the guarantor) to know whether the object can afford the communication and reputation to evaluate the trustworthiness and to detect the misbehaving node. Moreover, the guarantor is employed to find the object to get the service, and then reputation is employed to evaluate the trustworthiness of the selected object and to detect the misbehaving objects. The performance of the proposed model is carried by varying the malware probability (i.e., percentage of malicious objects). 

Chen \etal in \cite{doi:10.1002/dac.2930} delineate an energy-aware access scheme for service recommendation in SIoT that takes into consideration of the heterogeneous and decentralized environment. The model utilizes the reputation from experience, social relationships in terms of friendship and community of interest, and energy status to evaluate the trustworthiness of a node. This energy status consideration 
allows the balanced distribution of workload among the trustworthy nodes to improve the overall performance. Finally, the effectiveness of the proposed scheme is carried out in terms of rating accuracy, dynamic behaviour and network stability. The decentralized self-enforcement trust management model is presented by Azad \etal in \cite{8949450} that utilizes the weighted reputation through feedback generation to compute the trust of an object. The proposed model has three steps: 1) key generation through homomorphic encryption for privacy-preserving and post a public key to a bulletin board, 2) the generated public key 
is 
downloaded by objects, and 3) 
the reputation of objects is computed through weighted reputation. 
The self-enforcement is achieved via an automatic trust update through public verifiability by its peers in the network with zero knowledge proof. Finally, the performance of the model is carried out by taking into account the bandwidth required for committing feedback and communication overhead. 

A reputation and knowledge-based trust model is discussed 
in \cite{nguyen2016} wherein the reputation incorporates two trust metrics: recommendation and reputation, and knowledge assess an object and its respective owner to compute knowledge trust metrics of a service. To deal with the ambiguous knowledge with vague terms, i.e., ``good'', ``bad'', `'high'', and ``low'', a fuzzy logic-based mechanism is introduced to transform the human knowledge into object knowledge. Furthermore, a trust service platform is introduced that employs three components: \textit{trust agent}, \textit{trust broker} and \textit{trust management and analysis}. Trust agent is employed to collect the trust-related data in the SIoT domain, and trust broker is utilized to disseminate the trust-related data to numerous applications and services. Finally, the trust management and analysis component implements required task including but not limited to knowledge evaluation mechanisms, information model, reasoning mechanisms, and trust computation algorithm.

\begin{table*}[ht]
 \footnotesize
 \setlength{\arrayrulewidth}{.075em}
  \begin{center}
    \caption{Trust computation techniques}
    \label{tab:policy}
    \setlength{\leftmargini}{0.3cm}
   \begin{tabular}{|>{\centering\arraybackslash}m{1.75cm}| >{\centering\arraybackslash}m{2.9cm}| >{\centering\arraybackslash}m{5cm}| >{\centering\arraybackslash}m{5cm}| >{\centering\arraybackslash}m{1.25cm}|}
     \hline 
             \multicolumn{5}{c}{\text{\textbf{\normalsize Policy-based Techniques}}} \\
             \hline
             \rowcolor{Gray}
            \small \textbf{Ref.} & \small \textbf{Trust metrics} & \small \textbf{Strengths} & \small \textbf{Limitations} & \textbf{Context-awareness} \\ 
      \hline \hline
       \small Magarino \etal$\textbf{\cite{8468022}}$ & \begin{tableitems}[nosep,after=\strut]
                        \item Direct observation
                        \item Reputation
        \end{tableitems} 
        & \begin{tableitems}[nosep,after=\strut]
                        \item The framework presents a enhanced security mechanism by exploiting prioritization rules, certificates and trust management policies to detect hijacked nodes in the network. 
                        \item The performance of the model is improved when compared with the control approach that does not employ the trust and reputation mechanisms to detect the hijacked node. \vspace{-1em}  
                        \end{tableitems} 
        & \begin{tableitems}[nosep,after=\strut]
                        \item Performance evaluation of the model with many trust related attacks such as BMA, BSA, OOA, etc., is not known. \vspace{-1em} 
        \end{tableitems}  & No \\ \hline
        \small Al-Hamadi \etal$\textbf{\cite{8002573}}$ & 
        \begin{tableitems}[nosep,after=\strut]
                        \item Location rating
                        \item Raters trust
                        \item Witness trust
        \end{tableitems} 
        & \begin{tableitems}[nosep,after=\strut]
                        \item A trust-based decision making for IoT health system is proposed that considers risks, reliability trust, and health probability for decision making.
                        \item The model outperforms the traditional trust computational baseline protocol that filter out only the untrustworthy source without considering the reliability of the source and the relation between the user's health the level of harm. \vspace{-1em}    
        \end{tableitems} 
        & \begin{tableitems}[nosep,after=\strut]
                        \item Trust computational process of this model relies on static trust parameter, the optimal value for these parameters for dynamic environment is not known.  \vspace{-1em}  
        \end{tableitems}  & Yes\\ \hline
        \small Li \etal$\textbf{\cite{7959573}}$ & \begin{tableitems}[nosep,after=\strut]
                        \item Data observation
                        \item Experience as a history
        \end{tableitems} 
        & \begin{tableitems}[nosep,after=\strut]
                        \item A policy-based secure and trustworthy model is proposed to assess the trustworthiness of both the user and the data.
                        \item The model can efficiently detect three types of trust attacks (BMA, BSA, and OOA). \vspace{-1em} 
        \end{tableitems} 
        & \begin{tableitems}[nosep,after=\strut]
                        \item Policies are context dependent, and for highly dynamic IoT applications, system needs to update the policies very frequently or before/after every transaction. \vspace{-1em} 
        \end{tableitems}  & Yes\\ \hline
        \small Chen \etal $\textbf{\cite{10.1007/s11277-017-5120-4}}$ & \begin{tableitems}[nosep,after=\strut]
                        \item Energy status 
                        \item Bandwidth
                        \item Quality-of-provider
        \end{tableitems} 
        & \begin{tableitems}[nosep,after=\strut]
                        \item The model presents the trust management system exploiting the concept of maximum ratio combining (MRC) and source combining (SC) for IoT security protection. 
                        \item Performance evaluation of the model is carried out using QoS based trust score with and without MRC based trust. The MRC embedding with trust system performs well under different QoS level.  \vspace{-1em} 
            \end{tableitems} 
        & \begin{tableitems}[nosep,after=\strut]
                        \item The evaluation of the model is carried out on limited number of nodes that does not guarantee the scalability.   
                        \item Defence mechanism for different type of trust related attacks are is addressed. \vspace{-1em} 
        \end{tableitems}  & No \\ 
      \hline 
    \end{tabular}
  \end{center}
  \vspace{-3mm}
\end{table*}


Decisively, the reputation-based trust mechanisms have the upper hand while isolating the untrustworthy node for future endeavor but the inclusion of experience has its challenges such as how the old rating influences the current trust evaluation, how to include the forgetting factor for older rating as the characteristic of trust changes rapidly and it is important to include the recent rating, etc. The comparison of a number of schemes relying on a reputation-based trust model is 
presented 
in Table \ref{tab:reputation}.

\subsubsection{Prediction-based Trust Scheme (PredTS)}
The trust management system in prediction-based model takes into account the current and historical observation to identify and isolate the misbehaving node along with the improved trust computation process to overcome the limitation of trust aggregation techniques in particular the weighted sum approach \cite{8725509}. The prediction-based systems especially the machine-learning or deep learning approaches have upper hand when the trust composition step has more number of trust metrics in comparison with the weighted sum approach. The weighted sum approach fails to obtain the weights of each trust metric to get the single trust score as their can be infinite number of possibilities of assigning weights to each trust metric in different IoT environments \cite{8364607}. A number of prediction-based schemes \cite{8364607}\cite{10.1007/978-3-030-34986-8_28}\cite{9305298}\cite{9328463}\cite{sagar1}\cite{7986378}\cite{JAFARIAN2020107254}\cite{7986574} are described as follows and are compared in Table \ref{tab:prediction}.   

Jafarian \etal \cite{JAFARIAN2020107254} delineate a discrimination-aware trust model by taking into consideration of discriminatory behaviour of objects in SIoT network. An object's discriminatory behaviours can be attributed due to various reasons including but not limited to unavailability of computational resources and strong and weak ties of objects in terms of their social relationships. Furthermore, a weighted-KNN method is employed to ascertain the trust score by segregating the past and current rating of a an object (i.e., or service provider). The context-awareness is incorporated as a weight for each rating via a service rating query as rating vector $<S,E,SS,f>$ wherein $S$ represents service, $E$ is the energy status, $SS$ is the social similarity and $f$ is the feedback. Finally, the performance is analyzed in presence of numerous trust-related attacks by using the dataset from \cite{8580830}. 

A data-centric machine learning-based trust evaluation mode is proposed by Jayasinghe \etal in \cite{8364607} that incorporates the social trust metrics to evaluate the trustworthiness of nodes where the machine learning-based trust aggregation is used to get the single trust score. The machine learning-based aggregation has two steps of clustering (e.g., 
K-means) and classification (e.g., 
Support Vector Machine (SVM)) to identify the nodes as trustworthy or untrustworthy. Similarly, Marche \etal \cite{9305298} introduce a trust-related attack detection model for SIoT wherein the trust computation process involves two steps: \textit{training phase} and \textit{steady state phase}. In the training phase, trust is computed by employing the three trust metrics: 1) computation capability, a static characteristic of an object to distinguish the powerful devices, 2) relationship factor to consider the relationships between the objects, and 3) external opinion, to obtain the recommendations from neighbouring friends. Furthermore, training phase is utilized as an initial knowledge for the steady state phase. The steady-state step utilizes the initial dynamic knowledge to continuously learn the behaviour of object. To continuously learn the dynamic knowledge, an incremental SVM is employed with goodness, usefulness and perseverance score to quantify the trust of an object.  

A matrix factorization model is presented in \cite{9328463} where, at first, SIoT is demonstrated as bipartite graph in terms of service requester and service provider, then a hellinger distance is used to build a social network among service requester, and finally the matrix factorization is used to identify the trustworthy service provider. The model performs well under data sparsity, mitigates cold start problems, and is efficient in identifying malicious objects. Nevertheless, performance evaluation in presence of many trust-related attacks and the discussion on suitability of bipartite graph is not known. A social similarity-based trust computational model is presented in \cite{sagar1} where a k-means clustering and random forest classification is used to analyze the trust of the nodes over a period of time. Nevertheless, the proposed solution has no defence mechanism to tackle the trust attacks and is computationally expensive that leads to high latency in dynamic changing environments. 

Moreover, a deep learning model is delineated by Masmoudi \etal \cite{10.1007/978-3-030-34986-8_28} to segregate malicious nodes performing trust-related attacks, the trust computation process in this model follows a two-step process i.e., trust composition that includes social and QoS metrics, and deep learning-based trust aggregation. Nevertheless, a deep learning aggregation costs more computation power as well as increases the computation latency in dynamic environments. Abderrahim \etal \cite{7986378} present a trust management system employing community-of-interest based trust metrics and Kalman filter to predict the trustworthiness of nodes. The proposed system considers an on-off attack to evaluate the performance, however, it is equally important to prove the validity of the model in presence of other trust attacks. In general, the prediction-based scheme has the strength of providing a reliable trust aggregation to segregate the trust metrics and to make the precise trust decision, nonetheless, the computation cost of the prediction model particularly for machine and deep learning needs an optimal and low-cost solution.           

\begin{table*}[ht]
\footnotesize
\setlength{\arrayrulewidth}{.075em}
  \begin{center}
    \caption{Trust management components employed in SIoT}
    \label{tab:trust_components}
    \begin{tabular}{ |>{\centering\arraybackslash}m{1.775cm}| >{\centering\arraybackslash}m{1.8cm}| >{\centering\arraybackslash}m{2.175cm}| >{\centering\arraybackslash}m{1.55cm}| >{\centering\arraybackslash}m{1.45cm}| >{\centering\arraybackslash}m{1.8cm}| >{\centering\arraybackslash}m{1.9cm}| >{\centering\arraybackslash}m{2.15cm}|}
     \hline
     \rowcolor{Gray}
      \small\textbf{Ref.} & \small\textbf{Trust Metrics} & \small\textbf{Trust Aggregation} & \small\textbf{Trust Update} & \small\textbf{Trust Formation} & \small\textbf{Trust Propagation} & \small\textbf{Trust Decision} & \small\textbf{Trust Attacks}\\ 
      \hline \hline
      
      \small Nitti \etal $\textbf{\cite{Nitti2014}}$ & Social & Weighted Sum & Time-Driven & Multi-Trust & Distributed and Centralized & Threshold-based & SPA, WA, OSA, BMA, BSA \vspace{-0.8em}\\ \hline
      
      \small Truong \etal $\textbf{\cite{8254523}}$ & Social & Weighted Sum & Event-Driven & Multi-Trust & Centralized & Threshold-based & NA \\ \hline
      
      \small Khani \etal $\textbf{\cite{10.1007/978-3-030-03596-9_9}}$ & Social and QoS & Weighted Sum & Time-Driven & Multi-Trust & Distributed & Threshold-based & SPA, OOA, BMA, BSA \\ \hline
      
      \small Jayasinghe \etal $\textbf{\cite{8364607}}$ & Social & Machine Learning-based & Event-Driven & Multi-Trust & Distributed & Threshold-based & NA\\ \hline
      
      \small\ Chen \etal $\textbf{\cite{7097037}}$ & Social and QoS & Weighted Sum & Time-Driven & Multi-Trust & Distributed & Threshold-based & BMA \\ \hline
    
      \small Xia \etal $\textbf{\cite{8737491}}$ & Social & Fuzzy Logic & Event-Driven & Multi-Trust & Distributed & Threshold-based & SPA, OSA, OOA, BMA, BSA \\ \hline
    
      \small Xiao \etal $\textbf{\cite{7289151}}$ & Social and QoS & Weighted Sum and Bayesian System & Event and Time-Driven & Single-Trust & Distributed & Threshold-based & SPA, BMA, BSA \\ \hline
     
      \small Chen \etal $\textbf{\cite{doi:10.1002/dac.2930}}$ & Social & Weighted Sum & Event and Time-Driven & Multi-Trust & Distributed & Threshold-based & SPA, BMA, BSA \\ \hline
      
      \small Magarino \etal $\textbf{\cite{8468022}}$ & Social & Weighted Sum & Event-Driven & Multi-Trust & Centralised and Distributed & Context-based & NA\\ 
      \hline 
      
     
       \small Marche \etal $\textbf{\cite{9305298}}$ & Social and QoS & Machine Learning-based & Event-Driven & Multi-Trust & Distributed & Threshold-based & SPA, WA, OSA, OOA, BMA, BSA, DA\\ \hline 
     
      \small Al-Hamadi \etal$\textbf{\cite{8002573}}$ & Social and QoS & Weighted Sum & Event-Driven & Multi-Trust & Distributed & Threshold/Context-based & SPA, OSA\\ \hline

       \small Li \etal $\textbf{\cite{7959573}}$ & Social and QoS & Belief Theory & Event-Driven & Single-Trust & Distributed & Context-based & OOA, BMA, BSA\\ \hline
        
       \small Wei \etal $\textbf{\cite{9211717}}$ & Social and QoS & Weighted Sum & Event-Driven & Multi-Trust & Distributed & Threshold-based & OSA, BMA, BSA\\ \hline
      
        
       \small Azad \etal $\textbf{\cite{8949450}}$ & Social and QoS & Weighted Sum & Event-Driven & Multi-Trust & Distributed & Threshold-based & NA\\ \hline
        
        \small Truong \etal $\textbf{\cite{nguyen2016}}$ & Social and QoS & Fuzzy Logic & Event-Driven & Multi-Trust & Centralized & Threshold-based & NA\\ \hline

        
      \small Aalibagi \etal $\textbf{\cite{9328463}}$ & Social & Filtering & Even-Driven & Multi-Trust & Distributed & Threshold-based & OSA\\ \hline  
    
      \small Sagar \etal $\textbf{\cite{sagar1}}$ & Social & Machine Learning-based & Event-Driven & Multi-Trust & Centralised & Threshold-based & NA\\ \hline      

      \small Abderrahim \etal $\textbf{\cite{7986378}}$ & Social & Weighted Sum & Event-Driven & Multi-Trust & Distributed & Threshold-based & OOA\\ \hline 

      \small Chen \etal $\textbf{\cite{10.1007/s11277-017-5120-4}}$ & Social and QoS & Weighted Sum & Event-Driven & Multi-Trust & Distributed & Context-based & NA\\ \hline 
        
      \multicolumn{8}{|c|}{\text{\textbf{SPA} $\rightarrow$ Self-Promoting Attack, \textbf{WA} $\rightarrow$ Whitewashing Attack, \textbf{OSA} $\rightarrow$ Opportunistic Service Attack, \textbf{OOA} $\rightarrow$ On-Off Attack}}\\ 
      \multicolumn{8}{|c|}{\text{\textbf{BMA} $\rightarrow$ Bad Mouthing Attack, \textbf{BSA} $\rightarrow$ Ballot-Stuffing Attack, \textbf{DA (SBA)} $\rightarrow$ Discriminatory (Selective Behaviour) Attack, \textbf{NA} $\rightarrow$ No Attack}} \\\hline
    \end{tabular}
  \end{center}
  \vspace{-3mm}
\end{table*}

\begin{figure*}[ht]
    \centering
    \includegraphics[width=0.85\linewidth]{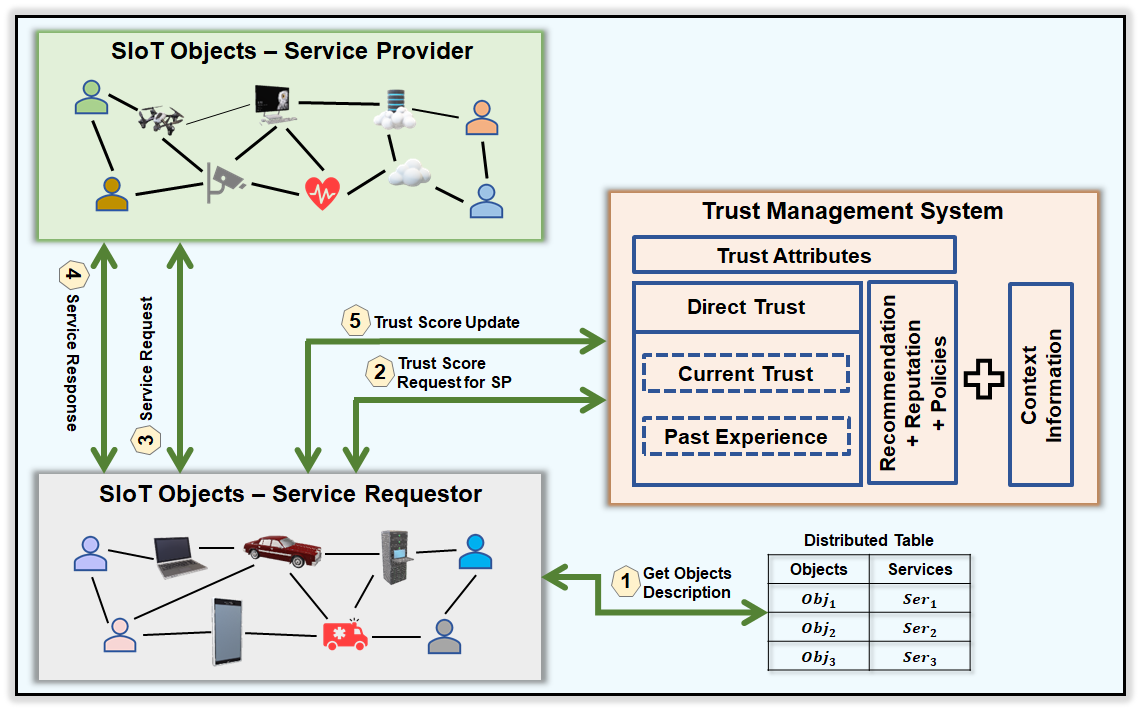}
     \caption{High-level overview of trustworthiness management system in service-oriented SIoT}
    \label{fig:trust_mang_model}
    \vspace{-1em}
\end{figure*}

\renewcommand\arraystretch{1.25}
\begin{table*}[ht]
	\small
\setlength{\arrayrulewidth}{.075em}
  \begin{center}
    \caption{Evaluation of trust management techniques using various dimensions}
    \label{tab:comp_const}
   \begin{tabular}{|>{\centering\arraybackslash}m{3.1cm}| >{\centering\arraybackslash}m{0.85cm}| >{\centering\arraybackslash}m{0.85cm}| >{\centering\arraybackslash}m{0.85cm}| >{\centering\arraybackslash}m{0.85cm}| >{\centering\arraybackslash}m{0.85cm}| >{\centering\arraybackslash}m{0.85cm}| >{\centering\arraybackslash}m{0.85cm}| >{\centering\arraybackslash}m{0.85cm}| >{\centering\arraybackslash}m{0.85cm}| >{\centering\arraybackslash}m{0.85cm}| >{\centering\arraybackslash}m{0.85cm}|}
     \hline 
       \small\textbf{Studies} & \textbf{\rot{Scheme}} & \small\textbf{\rot{Accuracy}} & \small \textbf{\rot{Adaptability}} & \small \textbf{\rot{Availability}} & \textbf{\rot{Integrity}} & \textbf{\rot{Reliability}} & \textbf{\rot{Privacy}} & \textbf{\rot{Scalability}} & \textbf{\rot{Credibility}} & \textbf{\rot{Applicability}} & \textbf{\rot{Response}} \\ 
      \hline \hline
      \small Nitti \etal \textbf{\cite{Nitti2014}} & RecTS & H
        & P & NA & NA & LR & NA & HS & NA & NSA & NEET \\ \hline
       \small Truong \etal \textbf{\cite{8254523}} & RepTS & NK
        & F & NA & NA & HR & NA & HS & NA & NSA & NEET \\ \hline
       \small Khani \etal \textbf{\cite{10.1007/978-3-030-03596-9_9}} & RecTS
        & H & P & NA & NA & HR & NA & HS & NC & NSA & NEET\\ \hline
        \small Jayasinghe \etal $\textbf{\cite{8364607}}$ & PredTS
        & H & F & L & H & NA & NA & LS & NA & NSA & EET\\ \hline
      \small  Chen \etal $\textbf{\cite{7097037}}$ & RecTS
        & H & F & L & NA & LR & PP & HS & FC & SA & NEET\\ \hline
      \small Xia \etal $\textbf{\cite{8737491}}$ & RecTS
        & H & F & NA & NA & HR & NA & HS & NA & NSA & NEET\\ \hline
      \small Xiao \etal $\textbf{\cite{7289151}}$ & RepTS
        & NK & NA & NA & NA & NA & NA & HS & NC & NSA & NEET\\ \hline
      \small Chen \etal $\textbf{\cite{doi:10.1002/dac.2930}}$ & RepTS
        & H & P & NA & NA & NA & NA & HS & NC & NSA & EET\\ \hline
      \small Magarino \etal $\textbf{\cite{8468022}}$ & PolTS
        & NK & F & H & NA & LR & NA & HS & FC & SA & EET\\ \hline
      \small  Marche \etal $\textbf{\cite{9305298}}$ & PredTS
        & H & F & H & NA & HR & NA & LS & NA & NSA & NEET\\ \hline
      \small Al-Hamadi \etal $\textbf{\cite{8002573}}$ & PolTS
        & H & F & H & L & HR & NA & HS & NA & SA & EET\\ \hline
     \small Li \etal $\textbf{\cite{7959573}}$ & PolTS
        & H & P & L & NA & LR & NA & LS & NC & SA & NEET\\ \hline
      \small Wei \etal $\textbf{\cite{9211717}}$ & RecTS
       & H & F & H & L & HR & NA & HS & NC & NSA & EET\\ \hline
      \small Azad \etal $\textbf{\cite{8949450}}$ & RepTS
        & H & F & NA & H & HR & PP & LS & NC & SA & EET\\ \hline
      \small Truong \etal $\textbf{\cite{nguyen2016}}$ & RepTS
        & NK & P & H & L & LR & NA & HS & NC & SA & NEET\\ \hline
     \small Aalibagi $\textbf{\cite{9328463}}$ & PredTS
        & H & F & H & H & HR & NA & HS & NC & NSA & EET\\ \hline
     \small Sagar \etal $\textbf{\cite{sagar1}}$ & PredTS
        & H & F & NA & H & NA & NA & LS & NA & NSA & EET\\ \hline
     \small Abderrahim \etal $\textbf{\cite{7986378}}$ & PredTS
        & NK & F & NA & NA & HR & NA & HS & NA & NSA & EET\\ \hline \small Chen \etal $\textbf{\cite{10.1007/s11277-017-5120-4}}$ & PolTS
        & L & P & L & NA & LR & NA & HS & NA & NSA & NEET\\ \hline
      \multicolumn{12}{|c|}{Trust Management Schemes} \\ \hline
      \multicolumn{12}{|c|}{\textbf{RecTS}: Recommendation-based Trust Schemes, \textbf{RepTS}: Recommendation-based Trust Schemes} \\
      \multicolumn{12}{|c|}{\textbf{PredTS}: Prediction-based Trust Schemes,  \textbf{PolTS}: Policy-based Trust Schemes} \\ \hline
      \multicolumn{3}{|c|}{Accuracy} &
      \multicolumn{3}{|c|}{Adaptability} &
      \multicolumn{3}{|c|}{Availability} &
      \multicolumn{3}{|c|}{Integrity}\\ \hline
      \multicolumn{3}{|c|}{H $\rightarrow$ High} &
      \multicolumn{3}{|c|}{F $\rightarrow$ Full} &
      \multicolumn{3}{|c|}{H $\rightarrow$ High} &
      \multicolumn{3}{|c|}{H $\rightarrow$ High}\\ 
      \multicolumn{3}{|c|}{L $\rightarrow$ Low} &
      \multicolumn{3}{|c|}{P $\rightarrow$ Partial} &
      \multicolumn{3}{|c|}{L $\rightarrow$ Low} &
      \multicolumn{3}{|c|}{L $\rightarrow$ Low}\\ 
      \multicolumn{3}{|c|}{NK $\rightarrow$ Not Known} &
      \multicolumn{3}{|c|}{NA $\rightarrow$ Not Addressed} &
      \multicolumn{3}{|c|}{NA $\rightarrow$ Not Addressed} &
      \multicolumn{3}{|c|}{NA $\rightarrow$ Not Addressed}\\ \hline
      \multicolumn{3}{|c|}{Reliability} &
      \multicolumn{3}{|c|}{Privacy} &
      \multicolumn{3}{|c|}{Scalability} &
      \multicolumn{3}{|c|}{Credibility}\\ \hline
      \multicolumn{3}{|c|}{HR $\rightarrow$ High Reliability} &
      \multicolumn{3}{|c|}{PP $\rightarrow$ Preserve Privacy} &
      \multicolumn{3}{|c|}{HS $\rightarrow$ Highly Scalable} &
      \multicolumn{3}{|c|}{FC $\rightarrow$ Feedback Credibility}\\ 
      \multicolumn{3}{|c|}{LR $\rightarrow$ Low Reliability} &
      \multicolumn{3}{|c|}{NA $\rightarrow$ Not Addressed} &
      \multicolumn{3}{|c|}{LS $\rightarrow$ Less Scalable} &
      \multicolumn{3}{|c|}{NC $\rightarrow$ Node's Credibility}\\ 
      \multicolumn{3}{|c|}{NA $\rightarrow$ Not Addressed} &
      \multicolumn{3}{|c|}{} &
      \multicolumn{3}{|c|}{} &
      \multicolumn{3}{|c|}{NA $\rightarrow$ Not Addressed}\\ \hline
      \multicolumn{6}{|c|}{Applicability} &
      \multicolumn{6}{|c|}{Response} \\ \hline
      \multicolumn{6}{|c|}{SA $\rightarrow$ Specific Application} &
      \multicolumn{6}{|c|}{EET $\rightarrow$ Emphasis on Evaluation Time}\\ 
      \multicolumn{6}{|c|}{NSA $\rightarrow$ No Specified Application} &
      \multicolumn{6}{|c|}{NEET $\rightarrow$ No Emphasis on Evaluation Time}\\ \hline
    \end{tabular}
  \end{center}
  \vspace{-3mm}
\end{table*}

\subsubsection{Policy-based Trust Scheme (PolTS)}
Policy-based trust models depend on pre-defined policies. Policies are the preset rules to evaluate the trustworthiness of nodes to detect malicious behaviour of nodes that have been compromised. These policies rely on network configuration as well as contextual information and can be expressed in mathematical or in language form \cite{7397227} \cite{cao2020policy}. 
A number of policy-based trust management schemes on IoT are present in research literature \cite{8468022}\cite{8002573}\cite{7959573}\cite{10.1007/s11277-017-5120-4}\cite{6821746}, however, these 
schemes are not yet employed in the SIoT. Therefore, we have selected the studies that utilize the social behaviour of objects in terms of social trust metrics. We have compared the selected policy-based trust scheme studies as given in Table \ref{tab:policy} and a brief description of each main research 
is described in this section. 

Al-Hamadi \etal \cite{8002573} 
present an adaptive trust-based decision making for IoT health systems that rely on different factors including location rating, raters, and witness trust to evaluate the trustworthiness of nodes to eliminate the nodes providing the misleading information. The proposed system takes into consideration a number of static trust parameters for trust computation, however, it is important to provide the optimal parameters for different IoT environment. Policy-based security and trustworthy model named $RealAlert$ is proposed by 
Li \etal \cite{7959573} to estimate the trustworthiness of a node as well as the data. The model presets the policies based on contextual information to detect the compromising nodes and misleading information by evaluating the model under different trust attacks. Nevertheless, policies of the proposed model are context-dependent that require human expertise to update the policies for highly dynamic IoT applications.

Moreover, a trust management model is proposed in \cite{10.1007/s11277-017-5120-4} that combines maximum ratio combining (MRC) and selection combining (SC) to ascertain the trustworthiness of nodes. The trust evaluation process starts with weighting the extracted parameters in the MRC step, subsequently, the output is then transferred to SC to obtain the final single trust score. The performance evaluation shows a promising result, however, the model is evaluated on a limited number of nodes that do not guarantee scalability, and no defence mechanism in presence of a trust attack is considered. Correspondingly, Magarino \etal \cite{8468022} present an enhanced security framework by employing prioritization rules, digital certificates, and trust and reputation policies to perceive a hijacked node providing deceptive information. The trust and reputation policies are direct interaction dependent and reputation is the recommendation of other nodes in the network. The performance evaluation shows that their approach is better at detecting the hijacked nodes than the other compared approaches. However, the evaluation against trust-related attacks is not illustrated. 
Overall, in general, the policy-based trust models are more suitable for 
an IoT application that does not have a dynamic nature such as no mobile nodes, the similar context in terms of location and time. 
Nonetheless, with the dynamic changing environment, it is more challenging to manage and update the policies for different contexts.  

Furthermore, Table \ref{tab:trust_components} presents the trust management components 
(see Figure~\ref{fig:trust_mang})
vis-à-vis their utilization in the selected schemes. As evident from the table, the research literature has developed some consensus on a number of these trust components (e.g., 
trust metrics, trust update, trust formation, trust propagation, and trust decision), and accordingly employed similar approaches.
Nevertheless, the trust aggregation component is evolving and researchers are exploiting other approaches, including but not limited to machine learning, fuzzy logic, and belief theory to handle the same. 

On a whole, SIoT is foreseen as a network of service providers and consumers (i.e., service-oriented SIoT) with enhanced service discovery and network navigability encompassing different social relationships to employ numerous applications and services, and trust is the indispensable factor to utilize these services in an unbiased and efficient manner. In light of the comparative analysis and discussion on different trust management schemes, a generalized high-level overview of a trustworthiness management system in SIoT 
is depicted 
in Figure \ref{fig:trust_mang_model}. The generalized trustworthiness management follows a total of five steps, wherein \emph{step1} provides the service requester access to a distributed table to facilitates which object (service provider) provides what service, \emph{step2} enables the service requester to request the trust score of the objects providing the requisite service from the trust management system, \emph{step3} lets the object request the service from the service provider possessing the highest trust score, and finally, in \emph{step:4}, once the service response from the service provider is received, the service requester updates the trust score in the trust management system.  

\subsection{Analysis of Trust Management Schemes}
\label{Sec:ATMS}
This section evaluates the trust management schemes discussed in Section~\ref{sec:TMS} with a set of dimensions. The selection of these dimensions is considered based on the highly dynamic and distributed nature of the SIoT network \cite{10.1145/2522968.2522980}\cite{8353121}\cite{8789637}. This section discusses the selected dimensions, and the evaluation of the schemes is provided in Table \ref{tab:comp_const}:

\begin{itemize}
    \item[-] \emph{Accuracy}: It refers to the degree of correctness of a trust assessment, which can be ascertained via a percentage of identification of untrustworthy or malicious nodes by employing the appropriate trust evaluation methods that work well under the high percentage of malicious nodes in the network \cite{10.1145/2522968.2522980}. 
    \item[-] \emph{Adaptability}: Owing to the dynamic nature of SIoT, trust evaluation framework must adapt to the changes in a different context, i.e., environmental conditions, temporal factor, and energy status. Furthermore, adaptability can also be observed in terms of variation in the trust parameters, i.e., which specific trust parameters have to be used in which context and weighting each parameter accordingly in a different context \cite{8353121}.  
    \item[-] \emph{Availability}: The availability signifies that the network services must be available even in the presence of malicious entities. One of the objectives of providing trustworthiness management is to ensure that the malicious entities in the network have a minimum effect on the provision of network services \cite{10.1145/2522968.2522980}.    
    \item[-] \emph{Integrity}: The network integrity implies that the content of message is protected during the transmission between two objects. An important component of trust computation is to share the feedback and recommendation among the objects so that it could also be employed for trust score computation purposes. Thus, integrity is essential to prevent the data from being modified without consent \cite{8789637}.
    \item[-] \emph{Reliability}: Reliability is the ability of a system to perform its functionality in an uninterrupted manner and error free without any failure for a particular period of time. In trust management, computation of trust and reputation from past experience can be seen as a reliable system \cite{8353121}.
    \item[-] \emph{Privacy}: The privacy of an object refers to the private and confidential information disclosure during the interaction with other objects in the trust management system. The private information can be personal or the activity information (e.g., 
    the information on with whom the object interacted and the services used by the same) \cite{8789637}.
    \item[-] \emph{Scalability}: This dimension is important given the dynamic and distributed nature of SIoT, 
    which is significant for a trust management system to be scalable. Moreover, with the increase in the number of objects, accessibility and inquiries to the trust assessment results also increase, thus the trust management must be able to handle the scalable nature of SIoT \cite{10.1145/2522968.2522980}\cite{8789637}.  
    \item[-] \emph{Credibility}: This dimension indicates the quality of information that makes the consumer trust the service provider. In the trust management system, 
    credibility can refer to
    the object’s credibility (i.e., service provider’s credibility) or the credibility of the feedback for trustworthy decision making (in the case of a system utilizing the feedback for trust computation)\cite{10.1145/2522968.2522980}. 
    \item[-] \emph{Applicability}: This dimension signifies the specific applications for which the trust management is designed and the ability of the system to be utilized for various applications and services \cite{10.1145/2522968.2522980}. 
    \item[-] \emph{Response}: Response refers to the response time a trust management system takes to provide the trust assessment result. It is essential for the trust management system to be prompt enough to handle many trust assessment inquiries, update the trustworthiness of an object, and propagate the trust results \cite{10.1145/2522968.2522980}.
\end{itemize}

\begin{figure*}[ht]
    \centering
    \includegraphics[width=0.85\linewidth]{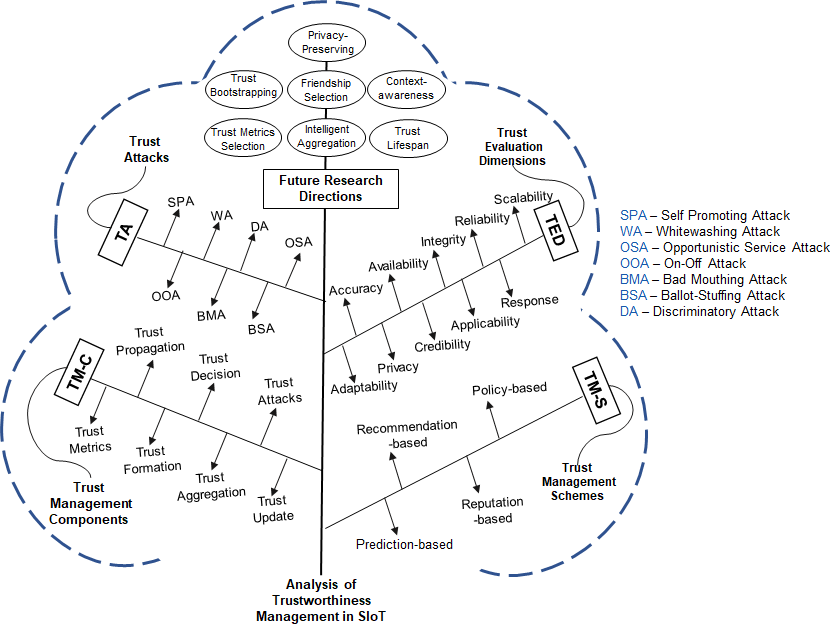}
     \caption{Summary of trustworthiness management analysis}
    \label{fig:trust_mang_sumnm}
    \vspace{-1em}
\end{figure*}

The evaluation of the trust management schemes vis-à-vis a set of dimensions is illustrated in Table \ref{tab:comp_const}. It can be 
observed 
that the recommendation-based schemes are highly accurate and scalable, nonetheless, have average performance in terms of adaptability, reliability, and applicability. Integrity, credibility, and availability remain the major concern in these schemes. Similarly, reputation-based schemes have higher accuracy, adaptability, and reliability, 
however, the performance of these schemes deteriorates in terms of integrity, availability, and credibility. The prediction-based schemes are fully adaptable and are highly accurate, nonetheless, they are not reliable, have low credibility, and are less scalable. Finally, it can be observed that the policy-based schemes are highly accurate, however, these schemes demonstrate major concerns in terms of adaptability, availability, reliability, and credibility. In general, the notion of privacy, credibility, integrity, and applicability in most of the schemes have not been addressed. 
They nonetheless, have laid the emphasis on the response of their proposed model. 

The overall discussion (and analysis) pertinent to the existing trust management schemes is illustrated in the form a ``tree'' (Figure \ref{fig:trust_mang_sumnm}), i.e., from categorizing the existing studies to the future research directions.

\section{Trust in SIoT-based Applications}
\label{sec:tsa}
The notion of SIoT can be utilized in numerous applications by employing social relationships among participant objects, and some of these applications are discussed in this section. 

\subsection{Crowdsourcing} 
Crowdsourcing focuses on the idea of outsourcing a task to a group of people for a business production \cite{Howe2006}. Recently, with the advancement in smartphones, and intelligent physical objects, crowdsourcing has emerged as an important platform for service-oriented IoT and is termed as \emph{IoT crowdsourcing} where IoT objects crowdsourced the services to other IoT objects. IoT objects with sensing and communication capabilities can crowdsource a wide range of applications and services including but not limited to \emph{computing resource} \cite{7214022} where a service provider can provide computing resources to low powered objects, \emph{ambient sensing} \cite{6516934} to sense the environment conditions, \emph{energy sharing} to provide wireless charging to the low energy level objects \cite{8487418}.
IoT crowdsourced can be more efficient by exploiting social relationships between service providers and service consumers by means of fast dissemination of information through the social network of objects \cite{8770228}. 

Recently, SIoT-enabled crowdsourcing on disaster reduction applications is proposed in \cite{Liu2019} wherein the Web-based map is designed to recruit the people and their devices (e.g., 
smartphones, tablets)
along with their social profile to massively transmit the disaster information to provide the disaster task force with enough information for relief support. With the advantage of providing numerous applications, crowdsourcing has its challenges, and providing trustworthy crowdsourcing is one of them wherein the system must guarantee the trustworthiness of crowdsourced services before relying on the information provided by them. Wang \etal in \cite{7721739} propose a trustworthy crowdsourcing model in SIoT to cope with the issue of trustworthiness of objects. The model considers two security aspects by encompassing a socially-aware message forwarding algorithm for social data link in SIoT, and a reputation-based mechanism to detect unreliable participants. Furthermore, a privacy-preserving incentive mechanism for crowdsourcing is proposed by Gian \etal in \cite{8770228} where the social relationships in terms of mutual friendship between computing entities are exploited for efficient utilization of resources and task completion. The inclusion of friendship between the workers in the large-scale SIoT not only benefits in obtaining help from friends but also suitable for handling collaborative tasks. 

In general, consumer-provider relationships enhanced the viability of crowdsourcing, however, many research challenges still need to be addressed, 
e.g., 
the trustworthiness of sensed data to prevent the use of polluted data, and trustworthiness of task computation results to counteract the invalid results from the dishonest participant trying to save their computing resources. 

\subsection{Smart Object Recommendation} 

In a service-oriented SIoT environment, an object can act as the service requester as well as the service provider and with billions of objects providing numerous services, it is significantly challenging to select the suitable objects providing the desired service, thus, the need for object recommendation and/or service recommendation appeared \cite{9167284}\cite{Sheng2021}. Similar to the recommendation systems in general, the service/object recommendation aims at suggesting the most relevant service to the requester. 

A framework for service recommendation in SIoT is proposed in \cite{9167284} wherein the social relationships among the participating objects are taken into consideration to provide the appropriate service recommendation. The employed object relationships are \emph{co-location}, \emph{co-work}, \emph{social}, \emph{co-owner}, and \emph{parental}. Furthermore, the boundary-based community detection algorithm is also proposed to detect the social communities among the objects and to enhance the service recommendation approach. Authors in \cite{Bok_2021} delineated user recommendation schemes for data sharing in SIoT by encompassing the interaction between the SIoT objects and the user. At first, the SIoT object preference is identified in terms of their interaction analysis with users. Subsequently, user interest keywords are extracted from users' social activities. Finally, the schemes recommend the top N users by analyzing the user similarity and SIoT object's preference.   

A time-aware smart object recommendation model for SIoT is presented in \cite{8936977} that encompasses the user's preference over a period of time and the social similarity of participating objects. Firstly, a latent probabilistic model is used to learn the user's preference in correspondence with their respective object's use. Secondly, object's social similarity is estimated by employing their social relationships. 
A recommendation list is then generated that utilizes the concept the item-based collaborative filtering. 

Overall, service/object recommendation will have a substantial impact on service-oriented SIoT systems. However, service/object recommendation has its own challenges including but not limited to selection of attributes and important relationship from social activity between the objects and among the users, how to include the relationships of highly mobile objects, how to protect the privacy of objects and users, and how to integrates the concept of context while recommending the service/object.

\subsection{Social Internet of Vehicles (SIoV)} 

The Internet of Vehicles (IoV) is the advancement in vehicular ad hoc networks (VANETs) and sensor networking techniques, and is conceptualized to solve numerous challenges, including but not limited to the lack of coordination among dissimilar vehicles traveling far from one another, information insufficiency, and scalability \cite{adnan2019}. SIoV is the modern trend of IoV \cite{nitti2014adding}, wherein social characteristics are integrated with the network of vehicles in a bid to offer 
new 
applications, 
e.g., 
personalized recommendation and route planning. 
In SIoV, a vehicle can socialize with other vehicles via sharing common interests, 
e.g., 
road situation, traffic information, weather conditions, and media sharing. Moreover, the social aspects in SIoV are not limited to vehicles only. In fact, they can include the socialization of drivers' and passengers' handheld devices, vehicular components, roadside units/infrastructures, etc \cite{BUTT201868}\cite{ATZORI2018132}. The implementation of SIoV is still in its infancy, nevertheless, a number of research articles have been published recently in terms of trust management \cite{lai2021}\cite{8527532}\cite{Gai2017}, computation offloading \cite{9356256} and other applications of SIoV (i.e., solution for traffic congestion, precise positioning, and vehicles' location protecting) \cite{9421372}\cite{9378811}\cite{9400866}.  

A trust-aware communication architecture for SIoV is proposed (\emph{TACASHI}) in \cite{8527532} comprising of five elements:  1) the vehicle, 2) vehicles' owners, 3) the passenger via his/her handheld devices, 4) roadside unit and other trust authorities, and 5) the online social network account of both drivers and passengers. Furthermore, the trust quantification process aggregates the intervehicular trust, roadside unit trust, location-related trust, and online social network trust. Moreover, the trust score may involve drivers' honesty based on their respective online social network profile. Similarly, Gai \etal \cite{Gai2017} delineate a reputation-based trust management model for SIoV, wherein each vehicle stores its reputation ascertained by other vehicles to avoid the loss of past transactions owing to a highly mobile network. Trust quantification involves multiple trust attributes aggregated together to ascertain a single trust score. The performance evaluation is carried out in terms of success rate and depicts high performance in presence of malicious vehicles. Nevertheless, the integrity of the model has not been discussed as a malicious vehicle possesses the potential to temper its past reputation to disrupt the functionality of a network. Furthermore, a friend matching scheme for SIoV has been proposed by Lai \etal \cite{lai2021} in an attempt to forbid the sensitive data leakage. The designed scheme is trust-based and ensure privacy preservation and can detect malicious vehicles and efficiently estimate vehicles' credibility to protect their privacy. The scheme encompasses three phases: 1) certificate issuance and update, i.e., a pseudonym is used as a vehicle identifier; 2) trust assessment, i.e., to estimate the credibility of the messages and accordingly, rate the respective vehicle, and 3) friend matching, i.e., by employing the trust scores of neighbouring vehicles having social relationships with each other and their corresponding certificates. The performance analysis is carried out in terms of network overhead and latency. Unfortunately, 
the performance in terms of malicious vehicles' eviction has not been considered.  

In general, the social relationships in SIoV have enhanced the viability of the IoV networks by facilitating the relationships between entities (e.g., vehicles, roadside units, and drivers' and passengers' handheld devices). These relationships are established by taking into consideration the context of mutual interest of the network entities and can be advantageous in several ways. For instance,  the transportation systems in smart cities can be further enriched with SIoV features by collecting the data from vehicles based on their social relationships and via taking smart decisions through intelligent analysis. Nevertheless, the nature of SIoV poses numerous research challenges, including but not limited to the highly dynamic nature of SIoV, managing social relationships of highly mobile entities, security, privacy and trust management, and lack of standard communication architecture.

\section{SIoT Simulation Tools and SIoT Datasets}
\label{sec:siot_data}
This section 
collects
the simulation tools utilized for SIoT and the datasets used for performance evaluation of SIoT-based models.  

\subsection{SIoT Simulation Tools}
With the extensive research in the emerging paradigm of SIoT, it is significant to identify the appropriate simulation tools that can be used to design the SIoT specific environment by integrating the social structure of objects. There are 
many simulation tools (e.g., 
OMNET++, NS-2, Cooja) that are utilized for the IoT environment \cite{10.1145/3109761.3158400}\cite{8234579}. However, not all of them are directly used for SIoT to address the complexity of the social structure of objects. This section highlights the simulation tools used for SIoT, especially for simulation and experimental analysis of trust management systems in SIoT. Some of the frequently used simulation tools used in the literature are discussed as follows:

\subsubsection{NetLogo} NetLogo is the open-source and a multi-agent programming module,
 which is suitable for natural as well as social phenomena \cite{netlogo}. With hundreds and thousands of independent agents, a researcher can give instructions to each one of these agents to explore and analyze the micro-level behaviour of objects/individuals from their interactions. Thus, it is appropriate for complex systems like SIoT. Most recently, this simulator along with the SWIM (\emph{Small World in Motion}) is used by many studies to evaluate the performance of their proposed trust management systems in SIoT \cite{Nitti2014}\cite{8737491}. SWIM is introduced as a mobility model for ad-hoc networking to generate the synthetic traces of mobility patterns to create a small world. Moreover, SWIM is also able to consider social behaviour similar to humans in real life and is statistically proven that the synthetic traces from SWIM are similar to that of humans \cite{5062134}. A few recommendation-based studies \cite{8737491}\cite{9211717} have utilized NetLogo simulator for experimental analysis of their work. 
 
\subsubsection{Network Simulator-3 (NS-3)} NS-3 is a discrete-event open-source simulator and is the successor of NS-2 \cite{riley2010ns}. 
It can be employed to create realistic simulation scenarios similar to real-world devices and protocols. Furthermore, NS-3 is documented as the popular tool for network simulation due to its flexibility, utilization in different fields and applications, adaptability to extend the resources for multiple application domains \cite{electronics9020272}. Overall, current literature suggests a number of studies on trust management have considered NS-3 simulator to validate their proposed model \cite{7097037}\cite{8002573}. 

\subsubsection{Objective Modular Network Testbed in C++ (OMNET++)} OMNET++ is another popular discrete event simulation tool extensively utilized in sensor networks research. Furthermore, OMNET++ is well-established and extensive, thus, it can integrate the external factor for specialized environment needs, 
e.g., 
to add the mobility for vehicular network \cite{veins}, incorporate the social profiles of objects to enhance the application capabilities \cite{desshpande2015}. In general, due to its flexible nature, this simulation tool can be utilized in various domains and applications.

\subsubsection{Others} There are numerous other well-established simulation tools 
that 
are considered in the literature for simulating the SIoT paradigm. Some of these tools are MATLAB, Python, Microsoft Visual Studio.
MATLAB is a popular multi-dimensional, multi-paradigm programming, and numerical computing platform utilized by many researchers to create models, develop algorithms, and analyze the data. Besides, MATLAB has a dedicated Simulink to design and deploy IoT applications and also offers flexibility and the possibility to integrate and analyze the data from third-party IoT service/platform (e.g., 
ThingSpeak \cite{thingspeak}). Similarly, research studies in SIoT have also considered Python as a simulation environment, especially for prediction-based studies. As a whole, MATLAB and Python have been the choice for many researchers to validate the performance evaluation of many trust managements system for SIoT \cite{8254523}\cite{10.1007/978-3-030-03596-9_9}\cite{8364607}\cite{sagar}\cite{9328463}\cite{sagar1}\cite{7986378}. There are several other least exploited simulation tools (e.g.,
GlomoSim \cite{10.1145/278009.278027}, Cooja \cite{10.1109/LCN.2004.38}) that are not commonly used by researchers in the literature.

\begin{figure}[!ht]
    \centering
    \includegraphics[width=0.8\linewidth]{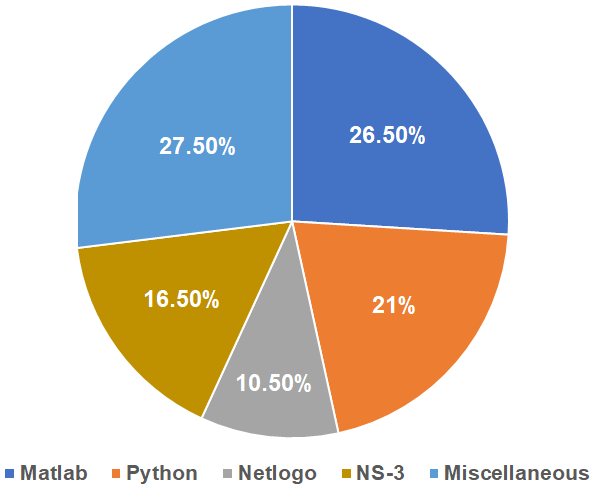}
     \caption{Percentage of simulation tools used in the selected articles}
    \label{fig:siot_simu}
    \vspace{-1em}
\end{figure}

Figure \ref{fig:siot_simu} shows the percentage of each simulation tool utilized for the evaluation of the trust management system in the research works discussed and analyzed in this survey. The miscellaneous part in the figure gives the percentage of articles where the name of simulation tools is not mentioned.

\subsection{SIoT Datasets} This section gives insight into the datasets currently present for evaluating the SIoT paradigm, especially, for trustworthiness management systems. Datasets are an important measure to evaluate and validate in an environment similar to real-world scenarios. Moreover, numerous datasets are available for IoT and social networks. However, these datasets can not be directly applied to the SIoT structure. Some of the datasets utilized in the literature are discussed as follows. 

The authors in \cite{marche2020exploit} collect the dataset that can be used to construct the SIoT Network. This dataset is based on real IoT objects employed in Santander city of Spain, which contains a total of 16,216 devices (14,600 for private users and 1,616 for the public service provider) with the description of each object in terms of $id\_device$ (Device Id), $user\_device$ (Owner Id), $device\_type$ (public or private device), $device\_brand$ (brand mapped in the form of number in range $1 \ to \ 12$), and $device\_model$ (models in range $1 \ to \ 24$). Furthermore, this dataset also includes the applications and services provided by each object, and the adjacency matrix providing the relationship (OOR, POR, CLOR, and SOR) between each object. In general, this dataset can be used to construct the SIoT network, nevertheless, validating the trust model is not possible as this dataset does not provide the interaction information between the device or any rating or reviews.

The frequently used dataset for evaluating the trust model in SIoT is the SIGCOMM-2009 dataset \cite{thlab-sigcomm2009-20120715}, 
which 
can be mapped in the form of a SIoT environment. This dataset contains the information of 76 objects in terms of their social profiles (friends and the communities they are involved in), and the interaction (15,776 number of interactions) between them. 
The dataset also provides the change in the objects' social profile with respect to time. In addition, researchers have utilized other popular datasets such as Epinions \cite{10.1007/978-3-540-39718-2_23} and Yelp\footnote{https://www.yelp.com/dataset} in combination with the SIoT dataset \cite{marche2020exploit} to integrate the social structure in order to validate the performance of their trust model. Epinions is the online social network consumer review site and used to decider whether to trust or each other or not, and contains more than 75,000 nodes and 500,000+ edges to describe the relationships. Finally, all the trust relationships interact and are combined with review ratings to show the reviews to the user. Similarly, Yelp is a form of social network where users can rate and review many businesses, which contains 1.6 million users, 6 million reviews, and 192,000 businesses. Besides, the Yelp dataset contains the user-user relationships. Due to limited real-world datasets for evaluation and validation, most of the researchers formulate their own datasets by taking into consideration the SIoT structure given in \cite{marche2020exploit}. Moreover, a few studies suggest the design of the testbeds to get the required dataset for performance evaluation \cite{8067499}\cite{10.1145/3384419.3430434}.

\section{Future Research Directions}
\label{sec:siot_future}
Although the notion of trust management in the context of SIoT has been widely explored and many noteworthy results have been 
proposed 
to date, there are still numerous research challenges that need 
attention of researchers.
This section
highlights the future research directions for trustworthiness management in SIoT.

\subsection{Trust Bootstrapping} 
Trust bootstrapping is also referred to as the \textit{cold start problem}. It is pertinent to note that the current trust management solutions presume the initial trust score of a newly joined SIoT object to be within the range $\{0, 0.5\}$. However, most of these solutions set the initial trust score of $0.5$ and classify the object as \textit{neutral} (i.e., neither \textit{trustworthy} nor \textit{untrustworthy}) \cite{7986378}\cite{7097037}. 
This assumption may lead a malicious SIoT object to jeopardize the basic functionality of a SIoT network before it is even identified as the untrustworthy object (or an object may perform the whitewashing attack where it changes its identity and joins the network with a new identity). Thus, it is essential to compute the initial trust of a newly enlisted SIoT object/device instead of using an arbitrary trust value. Recently, the authors in \cite{9328204} propose a trust framework for crowdsourced IoT services, wherein they utilize the social relationship among the owners of the devices to compute the initial relationship strength, the reputation of the device’s manufacturer as the initial reputation of that device, and the reputation of operating system that the device is using to avoid the limitation of presumed initial trust score. Nonetheless, the proposed solution still needs to assume that the reputation of the device is present and does not take into account the notion of social similarity between a public and a private device. Decisively, the combination of attributes, including but not limited to, social characteristics, long-term history, and reputation could be employed to get the initial trust score of a newly joined SIoT object.

\subsection{Friendship Selection} 
Friendship selection is an important factor since the service discovery in the SIoT paradigm is based on the relationship of an object with its friends in a bid to explore the friends of friends providing the specific service. These relationships are established, managed, and updated by an SIoT object and therefore, it is important to identify the right number of friends to prevent the resources (e.g., 
storage capacity) to be utilized for managing selfish objects. 
Selfish objects are referred to as the objects that intend to preserve their resources (e.g., 
energy and storage constraints), and utilize their resources for their own purpose or to enhance their reputation in the SIoT network. 
Furthermore, an imperative aspect of designing a trustworthiness management system for SIoT is to utilize the social attributes and these attributes exploit different types of relationships amongst the friends. Therefore, an efficient and appropriate friendship selection framework is required that is capable of employing different criteria to establish a number of relationships vis-à-vis different services. Moreover, the proposed framework should include a method to update the trustworthiness of existing as well as new friends to eliminate bad (e.g., selfish) friends. As of now, some possible strategies have been suggested by Nitti \etal \cite{6994231} for friendship selection, wherein an SIoT object sorts all of its friends in different order by their degrees (i.e., number of friends) to select the new friends in a bid to maximize its cluster and reachability in the whole network. One possible solution could be to maintain the interaction amongst the friends and the friends with maximum interactions within a specified duration should be added to the friendship list.  

\subsection{SIoT Specific Trust Metrics Selection} 
The key characteristic of SIoT is the integration of IoT and social networks. As of late, a number research studies consider hybrid SIoT trust metrics \cite{8949450}\cite{8885376}\cite{10.1007/s11277-017-5120-4}. In fact, the basic building block of a SIoT-based trust management system is the selection of appropriate trust metrics by taking into consideration application/service criteria primarily depending on dynamic environment (i.e., context information). 
Recently, 
a number of trust metrics are employed in some research studies \cite{9328463}\cite{9167284}, including but not limited to, similarity (e.g., 
friendship, community-of-interest, co-work, and co-location), cooperation between the SIoT objects (e.g., successful and unsuccessful interactions), recommendations, and reputation. However, it is not realistic to consider all the similarities for every application and service, as for a public service provider, it is not possible to ascertain the similarity score between the service consumer and the service provider. Therefore, the selection of trust metrics must follow an application's salient criteria and characteristic before designing an efficient trust management system.

\subsection{Context-awareness} 
Trust is a complex notion and varies with context (e.g., time, location, task, and energy status). In fact, each object trusts another object in different context \cite{10.1145/2501654.2501661}\cite{9264256}. Furthermore, owing to the dynamic nature of SIoT in terms of varied applications and services, the contextual information is important as the trust management system for a specific application and/or service may not be applicable for the other applications and services. A variety of context-aware trust models are proposed in the literature \cite{10.1007/978-3-030-03596-9_9}\cite{8725509} suggesting different contexts with the generally considered once being time, location, and objects' behaviours. However, some of the other contexts are equally important for an efficient trust management system. Therefore, it is important to design a trust model that considers not only the suitable trust metrics but also the context information in terms of where (i.e., location and environmental conditions), what (i.e., objects energy status and task), and when (i.e., temporal information) for the designed application. 

\subsection{Intelligent Trust Aggregation} 
Trust aggregation is an important component of trust management, wherein the selected trust metrics are aggregated to ascertain a single trust score. The conventional aggregation methods suggested in the literature \cite{Nitti2014}\cite{10.1007/978-3-030-03596-9_9} employ a linear weighted sum mechanism with randomly assigned weights, which can be either static or dynamic for each of the trust metrics. Nevertheless, the weighted sum approach has 
some disadvantages, including but not limited to, an infinite number of conceivable outcomes with regards to assessing a weighting factor for each metric and inability to recognize which trust metric makes the most impact on the overall trust in a specific environment. Consequently, there is a need of intelligent trust aggregation mechanism to overcome the limitations of conventional aggregation techniques. Lately, the idea of machine learning-based aggregation has been suggested by the researchers to obtain the weights of each metric in terms of its importance \cite{sagar}. However, machine learning-based solutions have their own limitations, e.g., 
these solutions are computationally expensive and results in increasing the computational latency. One possible solution to overcome these limitations is to design an optimized machine learning-based aggregation that aggregates the trust metrics of clusters of objects instead of all the objects in the network to train the models. 

\subsection{Trust Lifespan - Decay} 
It is evident that the trust of an SIoT object towards another object varies with time, however, these variations are subject to decay if there are no or neutral interactions between the objects \cite{8254523}\cite{9025055}. Owing to the SIoT intrinsic characteristic, SIoT objects during interactions may encounter many other objects, and therefore, it is not viable to store the trust of all the nodes from the past. 
It is imperative to consider the trust lifespan wherein trust score of inactive SIoT objects must be subject to decay after a particular duration of time. Truong \etal \cite{8254523} propose an experience-reputation model that gives the idea of trust decay over a period of time, wherein the trust of SIoT objects decline based on strong and weak tie (strong tie represents the strong relationship) with the other SIoT objects. However, the model does not discuss about the type of relationships required for ascertaining strong and weak ties. In the SIoT paradigm, different social relationships along with the number of interactions could be utilized to manage the trust lifespan.

\subsection{Privacy-preservation} 
It is pertinent to note that an adversary can eavesdrop on the private social profile of the owners of the objects and find the associated detail of the owners using online social networks. Hence, the privacy-preserving solutions are essential to address the risks involved and to promote the SIoT applications and services. Moreover, there are a few notable studies in the literature pertinent to privacy-preservation for trust management in SIoT \cite{7097037}\cite{8949450}. Chen \etal \cite{7097037} utilize the one-way hash function to encrypt the social information of nodes during the interaction, whereas Azad \etal \cite{8949450} use homomorphic encryption to protect the privacy of SIoT objects. Nevertheless, with only a few studies on privacy-preservation in SIoT, a novel and optimal framework of privacy-preserving is considered as an indispensable future research direction for trustworthy SIoT.  

\section{Conclusion}
\label{sec:conc}
Recently, the emerging paradigm of Social Internet of Things (SIoT) has become a vibrant and rapidly growing area of research. Trust is considered as the impediment for the adoption of social characteristics amongst the smart objects for establishing trustworthy social relationships and to provide reliable services. In this survey, we have presented a comprehensive discussion on trustworthiness management in SIoT. At first, we classify the trustworthiness techniques into four broad categories and the strengths and limitations of the referred studies under each of these categories are analyzed and compared. We further compare the referred studies in terms of trust management components and a set of assessment dimensions. Finally, we provide a high-level overview of the generic trust management framework for service-oriented SIoT, and put forward the future research directions to address various trust-related SIoT research issues.  

\section*{Acknowledgements}
Subhash Sagar's research is supported by the Higher Education Commission of the Government of Pakistan and the Macquarie University, Australia (Macquarie University's Research Excellence Award). Adnan Mahmood's research is funded via the Macquarie University's Postdoctoral Research Fellowship. Quan Z. Sheng's research is partially supported by Australian Research Council
(ARC) Discovery
Project DP20010229 and Linkage Infrastructure, Equipment and Facilities Projects LE180100158  and LE220100078. 


\bibliographystyle{IEEEtran}
\bibliography{sagar}

\begin{IEEEbiography}[{\includegraphics[width=1.05in,height=1.15in,clip,keepaspectratio]{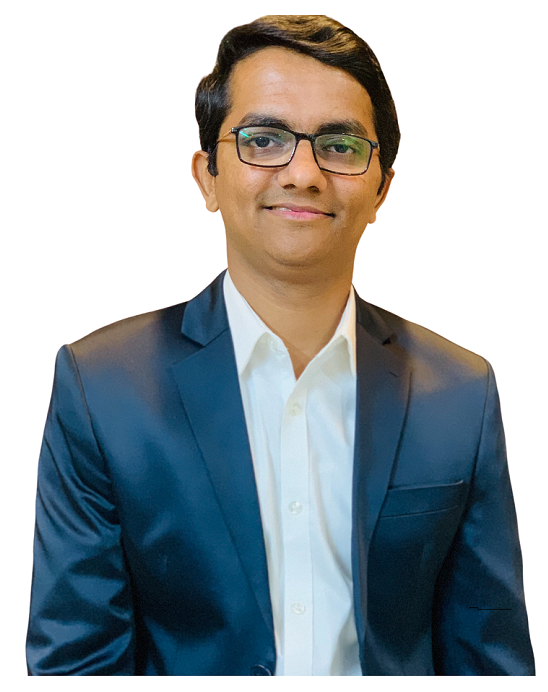}}]{Subhash Sagar}
received the BS in Electrical (Telecommunication) from the COMSATS Institute of Information Technology, Islamabad, Pakistan and a Master's degree in Computer Science from South Asian University, New Delhi, India in 2012 and 2016 respectively.
He is currently pursuing PhD degree at the School of Computing, Macquarie University, Sydney, Australia. Before moving to Macquarie University, Subhash worked as a faculty member at the Department of Computer Science, National University of Computer and Emerging Sciences, Karachi, Pakistan from 2017 to 2019. His current research interests include the Internet of Things, Social Internet of Things, and Trust Management.
\end{IEEEbiography}\vspace{-2mm}

\begin{IEEEbiography}[{\includegraphics[width=1in,height=1.25in,clip,keepaspectratio]{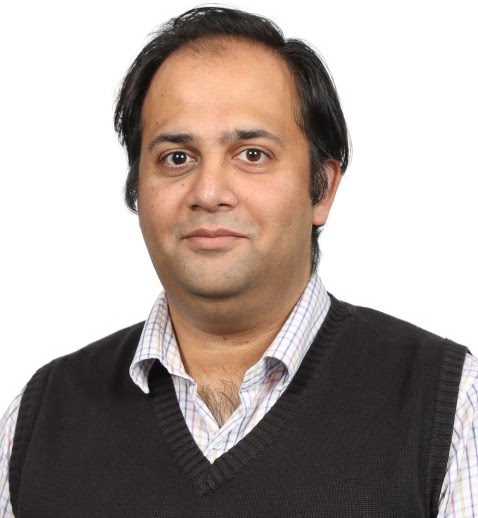}}]{Adnan Mahmood}
holds a PhD degree in Computer Science and is currently a Postdoctoral Research Fellow at the School of Computing, Macquarie University, Sydney, Australia. Before moving to Macquarie University, Adnan has spent a considerable number of years in the academic and research settings of Republic of Ireland, South Korea, Malaysia, Pakistan, and People’s Republic of China. His research interests include Software-Defined Networks, Intelligent Transportation Systems, Internet of Things (primarily the Internet of Vehicles), Trust Management, and the Next Generation Heterogeneous Wireless Networks. Adnan besides serve on the Technical Program Committees of a number of reputed International Conferences. He is a member of the IEEE, IET, and the ACM. 
\end{IEEEbiography}\vspace{-2em}

\begin{IEEEbiography}[{\includegraphics[width=1in,height=1.25in,clip,keepaspectratio]{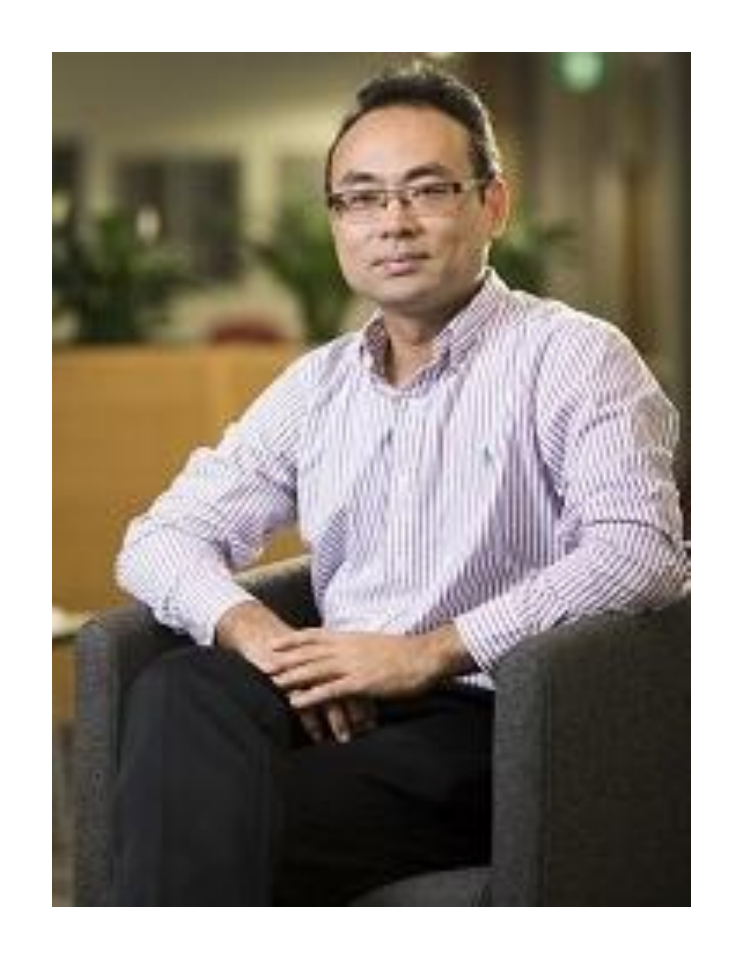}}]{Quan Z. (Michael) Sheng}
is a full Professor and Head of School of Computing at Macquarie University. Before moving to Macquarie, Michael spent 10 years at School of Computer Science, the University of Adelaide (UoA). Michael holds a PhD degree in Computer Science from the University of New South Wales (UNSW) and did his post-doc as a research scientist at CSIRO ICT Centre. From 1999 to 2001, Sheng also worked at UNSW as a visiting research fellow. Prior to that, he spent 6 years as a senior software engineer in industries. Prof Sheng's research interests include the Internet of Things, big data analytics, service computing, and Internet technologies. 
Dr. Michael Sheng is the recipient of the ARC Future Fellowship (2014), Chris Wallace Award for Outstanding Research Contribution (2012), and Microsoft Research Fellowship (2003). He is ranked by Microsoft Academic as one of the Most Impactful Authors in Services Computing (ranked top 5 all time). He is a member of the IEEE and the ACM.
\end{IEEEbiography}\vspace{-13mm}

\begin{IEEEbiography}[{\includegraphics[width=1in,height=1.25in,clip,keepaspectratio]{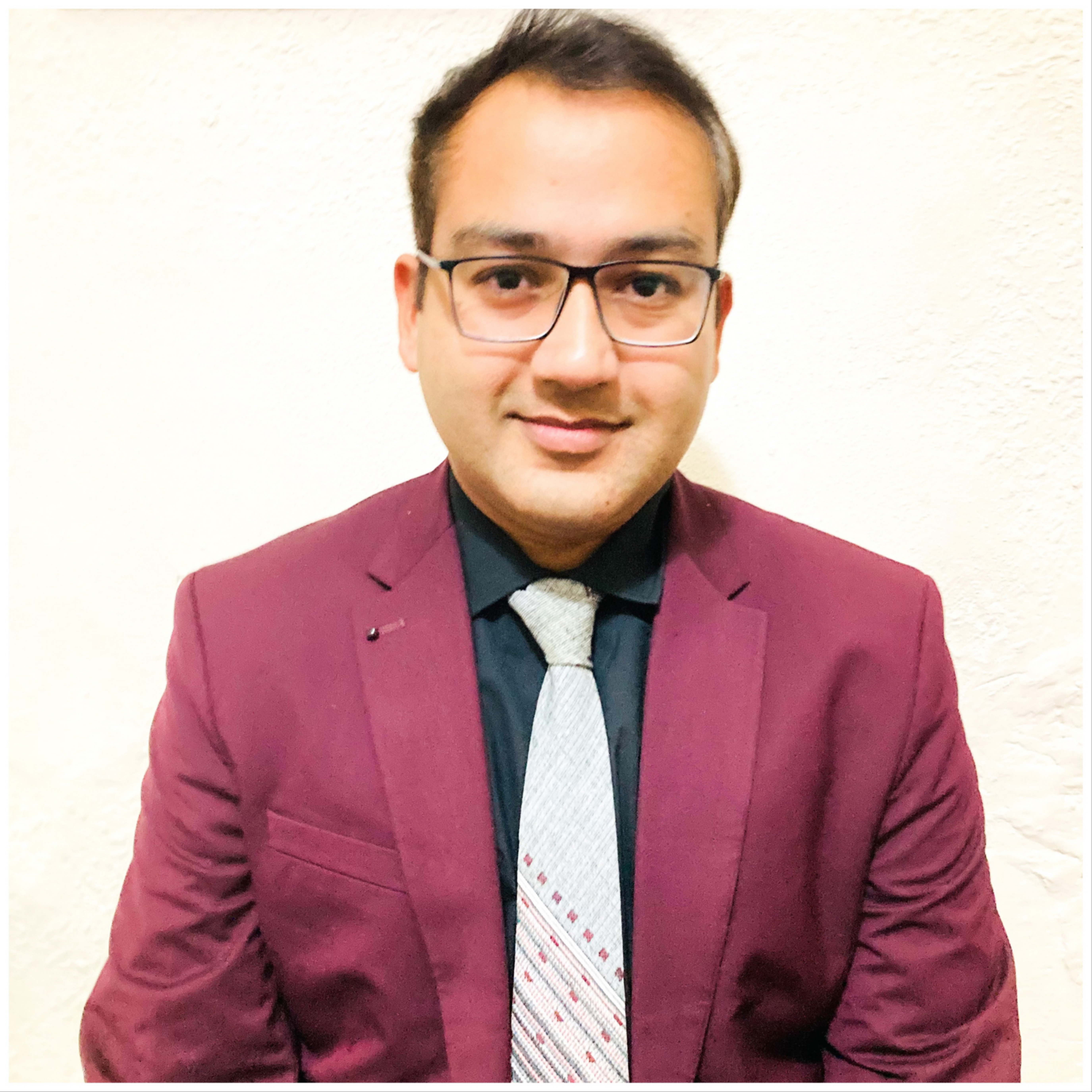}}]{Jitander Kumar Pabani}
received his BE degree in Telecommunication Engineering from Mehran University of Engineering and Technology Jamshoro, Sindh Pakistan, and Master of Engineering in Telecommunication Engineering from Hamdard University, Karachi, Pakistan in 2011 and 2014 respectively. He also completed his Post-Graduate Diploma in Statistics in 2017 from the University of Karachi, Pakistan. Currently, he is pursuing his PhD studies at the Department of Communication Engineering, Universidad de Malaga, Spain, funded through the Faculty Development Program by Dawood University of Engineering and Technology, Karachi, and Higher Education Commission of Pakistan. He has been also associated with Dawood University of Engineering and Technology in the capacity of Lecturer since 2016. He has more than 10 years of teaching experience. His area of research includes Underwater Wireless Sensor Networks, Wireless Body Area Networks, Internet of Things, Machine Learning, and Fuzzy Decision Making.
\end{IEEEbiography}\vspace{-13mm}

\begin{IEEEbiography}[{\includegraphics[width=1in,height=1.25in,clip,keepaspectratio]{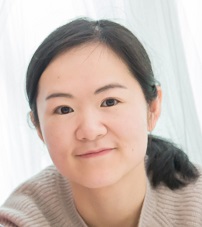}}]{Wei (Emma) Zhang}
is currently a Lecturer at the School of Computer Science, The University of Adelaide. She obtained her PhD in 2017 from the School of Computer Science, The University of Adelaide. Her research interests include text mining, deep learning, natural language processing, information retrieval, and Internet of Things (IoT)
applications. She has close to 100 publications to date as edited books and proceedings, refereed book chapters, and refereed technical papers in journals and conferences including ACM Computing Surveys, TOIT, ACM TIST, WWWJ, Communications of the ACM, ACL, SIGIR, WWW, EDBT, CIKM, ICSOC and CAiSE. Her PhD thesis has been published by Springer as a monograph. She is a member of the IEEE, the ACM and the ACL.
\end{IEEEbiography}

\end{document}